\documentclass[twocolumn,
floatfix,
nofootinbib,
showpacs,prb,aps]{revtex4-2}

\usepackage{dcolumn}
\usepackage{graphicx}	
\usepackage{amsmath}    
\usepackage{color}      
\usepackage{amsfonts}
\usepackage{amssymb}
\usepackage{enumitem}  
\usepackage{yfonts}
\usepackage{fourier}

\usepackage[%
          unicode=true,%
          colorlinks=true,%
          linkcolor=blue,anchorcolor=blue,%
          breaklinks=true,%
          citecolor=blue,filecolor=cyan,%
          menucolor=blue,%
          urlcolor=blue]{hyperref}
          
\def\i{\text{i}}

\def\e{\text{e}}
\def\d{\text{d}}

\def\be{\begin{equation}}
\def\ee{\end{equation}}
\def\bse{\begin{subequations}}
\def\ese{\end{subequations}}
\def\ba{\begin{align}}
\def\ea{\end{align}}

\def\ce{{\cal E}}
\def\pix#1#2{\includegraphics[#1]{#2}} 

\begin{document}

\title{Dynamics of a periodic $\boldsymbol{XY}$ chain coupled to a photon mode}
\author{S. Varbev}
\affiliation{Institute of Solid State Physics, Bulgarian Academy of Sciences, Tzarigradsko chauss\'{e}e 72, 1784 Sofia, Bulgaria}
\author{I. Boradjiev}
\email[boradjiev@issp.bas.bg]{}
\affiliation{Institute of Solid State Physics, Bulgarian Academy of Sciences, Tzarigradsko chauss\'{e}e 72, 1784 Sofia, Bulgaria}
\author{H. Tonchev}
\affiliation{Institute of Solid State Physics, Bulgarian Academy of Sciences, Tzarigradsko chauss\'{e}e 72, 1784 Sofia, Bulgaria}
\author{H. Chamati}
\affiliation{Institute of Solid State Physics, Bulgarian Academy of Sciences, Tzarigradsko chauss\'{e}e 72, 1784 Sofia, Bulgaria}

\date{\today}

\begin{abstract}

We study the real-time dynamics of a periodic $XY$ system exposed to a composite field
comprised of a constant homogeneous magnetic and
a quantized circularly polarized electromagnetic fields.
The interaction between the quantized mode and spin-magnetic moments is 
modeled by the Dicke Hamiltonian.  
The rotating wave approximation is applied and the 
conditions for its validity are discussed.
It is shown that
if initially all of 
the excitations are contained in the field, then in the regime of large detuning, 
the main evolutionary effect involves oscillations of the excitations between 
the zero-momentum modes of the chain and the field. 
Accordingly, the reduced photon number and magnetization 
per site reveal a sort of oscillatory behavior. 
Effective Hamiltonians describing the short-time dynamics 
of the present model for small number of excitations and large detuning are introduced.
The resonance case is considered in the context of photon emission from the chain 
initially prepared in the (partially) excited state. 
In particular, it is demonstrated, in the framework of a specific
example, that 
the superradiant behavior
shows up at the beginning of the emission, 
when we have an initial state
with a maximally excited $XY$ chain.
Possible applications of the model to problems such as spin chain and $J$-aggregate 
in a single-mode cavity are discussed.

\end{abstract}


\maketitle

\section{Introduction}
\label{Intro}

Light-matter interaction is an important mean in condensed matter
physics. It provides useful insights into the material's behavior and
may be used to manipulate its physical properties offering an extensive capability to engineer devices for a wide
variety of applications. In particular, the manipulation of
magnetic ordering and thus the magnetic properties of spin systems
through
light-matter interaction attracts
an ever
increasing interest due to potential applications in spintronics and
quantum information, see e.g.
\cite{kirilyuk_ultrafast_2010,noh_quantum_2017,harder_cavity_2018} and
references therein.
A commonly used experimental approach
consists in using focused ultra-short laser pulses
\cite{Beaurepaire_1995,kirilyuk_ultrafast_2010,bigot_ultrafast_2013, kimel_femtosecond_2007,zhang_all-optical_2016, zhang_ultrafast_2014}
to control the dynamics in magnetic systems,
such as, ferromagnetic, antiferromagnetic
\cite{bossini_controlling_2014} or ferrimagnetic materials
\cite{hansteen_femtosecond_2005}. Thus, a
single ultrafast laser pulse can
permanently affect the dynamics of a spin even in the absence of a magnetic field
\cite{zhang_all-optical_2016,zhang_ultrafast_2014}.
This accomplishment would not have been possible without the
technological advancement that made the synthesis of low dimensional
spin systems achievable \cite{brune_assembly_2006}, on one hand and
a better understanding of light-matter interaction to bring new tools for manipulating
quantum states -- a key ingredient for the ultimate goal of quantum
information -- the quantum computer, on the other.

In the majority of the
above mentioned studies, the control of spin systems by light
involves indirect interactions,
i.e. those relying mainly on the electric
component of the electromagnetic field only.
These could include for instance the inverse Faraday effect 
or other photomagnetic, optomagnetic or thermal effects \cite{kirilyuk_ultrafast_2010}.
Direct interactions between
spin-magnetic moments and the magnetic component of the
electromagnetic field are also present. 
Due to the small coupling, such an effect is (especially for optical frequencies) 
usually obscured by the indirect  
effects, and therefore such interaction mechanism is
negligible and is rarely taken into account. 
For a spin system containing a large number of magnetic
moments, 
when indirect effects are screened
or small enough, 
the direct interaction may be important since the coupling grows with the number of spins.
This assumption is used in Ref.
\cite{Chudnovsky2002} where the authors consider the
interaction of spin-magnetic moment with photon
and the possibility for superradiance from
crystals of molecular nanomagnets.
With respect to the driving electromagnetic field, 
one may also consider the number or
coherent states, instead of 
the quantum state of the laser light. 
These states may, in principle, be prepared in a cavity.
Interactions between a photonic cavity mode
and spins in a nanomagnet
turn out to produce a strong magnet-photon coupling \cite{Soykal2010}.

In quantum optics, the basic models describing a system consisting of two-level atom(s) 
interacting with a single mode of quantized electromagnetic field are: 
The Jaynes-Cummings \cite{Jaynes1963}, the Tavis-Cummings \cite{Tavis1968}, 
and the Dicke \cite{Dicke1954} models. 
The Dicke model describes a set of independent two-level systems 
interacting with a photon mode. 
Within the Rotating Wave Approximation (RWA) the Dicke model reduces
to the Tavis-Cummings model, 
which in the case of a single atom turns out to be the Jaynes-Cummings model. 
Over the years, these models have been and are continuously studied and 
extended in a great many directions, such as
transient and steady-state superradiance \cite{Dicke1954,kirton_introduction_2019}, superabsorption \cite{Higgins_superabsorption_2014}, 
the validity of RWA \cite{grinberg_beyond_2010,Agarwal2012}, 
the high detuning limit \cite{agarwal_atomic_1997,Klimov1998}, 
the inhomogeneous versions of the models \cite{Strater2012}, etc.
When switching on the interaction among the constituting atomic spins in
the Dicke (Tavis-Cummings) model it is possible to end up with a physically
rich system, where light and matter interact \cite{Tokihiro1993,Wu2016,Wu2018,tonchev_interaction_2016,tonchev_energy_2016,tonchev_energy_2019}.
Depending on the specificity of the problem different ways
can be used to (effectively) model the spin-spin interaction. A widely used
spin model is the exactly soluble quantum $XY$ model with nearest-neighbor interaction
\cite{Lieb1961a,E.Lieb1966b,DePasquale2008}.

In the present paper we consider a spin-magnetic system (in
order to achieve admissible coupling strength) directly coupled to a
single cavity mode.
We study the behavior of the quantum $XY$ spin chain subject to 
a constant homogeneous magnetic field and 
to a single mode of quantized electromagnetic field. 
The respective interaction is described by the Zeeman term 
which couples the spin-magnetic moments 
to the magnetic fields.
Instead of a spin chain one can think of a $XY$ system of interacting 
two-level atoms and a Stark term describing the interaction 
between the atomic electric dipole moments 
and the electric component of the mode. 
Then the formal Hamiltonian does not change, 
which makes the obtained results applicable to both types of problems. 
Here, we do not account for decoherence mechanisms, that is we deal with a closed system 
that does not interact with the outside world.
Hamiltonian of our system may be written down as the sum of the quantized 
electromagnetic field, the $XY$ model and the Dicke Hamiltonian.
Applying the Rotating Wave Approximation to the Dicke Hamiltonian we
explored the dynamics of the model depending upon the relevant
parameters. 
The model under consideration may also find applications in problems related to
a bosonic mode in a spin-bath, spin chains or linear molecular aggregates 
in a single-mode cavity, and in the study of superradiant and superabsorbing systems.

The rest of the paper is organized as follows.
We proceed with the description of the model in Sect. \ref{Sec_Model}. 
Sec. \ref{Sec_RWA} is devoted to the conditions for validity and application of RWA. 
In Sec. \ref{Sec_HinXYBasis} the model Hamiltonian is transformed to the fermion basis 
diagonalizing the $XY$ chain part. 
Some notations and definitions are given in Sec. \ref{Sec_NotationsAndDefinitions}. 
In Sec. \ref{Sec_SingleExcitation} the exact analytical solution for the 
single excitation case is derived. 
Sec. \ref{Sec_GeneralCase} deals with the formal solution for the general 
(multiple excitation) case. 
Also, the results from numerical simulations in the large detuning regime, 
given that at the beginning all the excitations are in the field, are presented. 
In Sec. \ref{Sec_EffectiveHamiltonians} effective Hamiltonians 
approximating the Hamiltonian under consideration are obtained.
In Sec. \ref{Sec_PhotonEmission} the evolution of the system, when all the excitations 
are contained in the spin chain at the beginning, is considered. 
Sec. \ref{Sec_PossibleApplications} presents
some potential applications of the model 
and the respective estimates of the energies.
The summary of the results is given in Sec. \ref{Sec_Conclusions}. 
There are two appendices:
In Appendix \ref{RWA} we present detailed
calculations of the conditions of validity of the RWA. 
In Appendix \ref{XY_Diagonalization} we recall the diagonalization procedure for the $XY$ 
spin chain model, discuss its ground state, 
and derive the full Hamiltonian in the
associated basis.

\section{The model}\label{Sec_Model}

The Hamiltonian of a chain of interacting (effective) spins, 
placed in a complex magnetic field consisting of cavity modes and a 
constant homogeneous component, may be written as the sum of three
terms, as follows
\begin{align}
\hat{H} &= \hat{H}^{CF} + \hat{H}^{SM} + \hat{V}.
\label{H_general}
\end{align}
Within this setup, we will assume an ideal cavity and
neglect any effect of the electric component of the field.

The first term of \eqref{H_general}, $\hat{H}^{CF}$
represents the quantized cavity field. It reads
\begin{align}
\hat{H}^{CF} &= \hbar\omega\left(\hat{n}_{x} + \hat{n}_y  \right),
\end{align}
where $\hat{n}_{x}$ and $\hat{n}_{y}$ are the number operators corresponding to the 
$x$ and $y$ polarized components of the magnetic field, 
and $\omega$ stands for their frequencies. 
Here and below, we skip the zero point energy.

The second term of Hamiltonian \eqref{H_general},
describing the interacting 1D spin
system is given by the $XY$ spin model \cite{Lieb1961a,E.Lieb1966b}
\begin{align}
\hat{H}^{SM}\equiv\hat{H}^{XY} = 2J \sum_{i > j} \left(\hat{S}_{ix} \hat{S}_{jx} + \hat{S}_{iy} \hat{S}_{jy}\right),
\end{align}
where $J$ is the coupling constant and $\hat{S}_i$ are dimensionless
spin-$\tfrac12$ (pseudo-)vector operators, whose components obey the following commutation relations
\begin{align}
[\hat{S}_{i\eta},\hat{S}_{j\zeta}] = \i\delta_{ij}\varepsilon_{\eta,\zeta,\theta} \, \hat{S}_{\theta} ,
\quad \eta,\zeta,\theta  = x,y,z,
\end{align}
with $i, \, j$ labeling the spin sites.

The remaining term of \eqref{H_general} is given by
\begin{equation}
\hat{V} = - \sum_{i}\hat{\mathbf{\mu}}_{S_i} \cdot
	\hat{\mathbf{B}},
\end{equation}
where we define the magnetic field $\hat{\mathbf{B}}$, and the
interaction between $\hat{\mathbf{B}}$ and the magnetic moments
$\hat{\mathbf{\mu}}_{S_i}$ associated with the spins in the $XY$
model.

\begin{figure}[h!]
	\pix{width=\columnwidth}{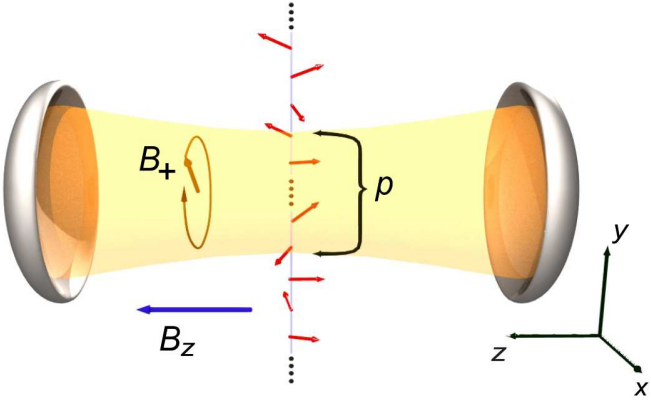}
	\caption{(Color online)
		Setup of the model. 
		$B_{+}$ and $B_{z}$ are the left circularly polarized and constant components 
		of the magnetic field respectively and the tilted arrows 
		represent the spin chain. 
		The right circularly polarized component of the field is not shown in the figure 
		since we discard it by means of the RWA, see Sec. \ref{Sec_RWA}.}
	\label{Fig:SystemGeometry}
\end{figure}

To proceed further, we use the relation between
the magnetic moments and the spins, i.e. $\hat{\mathbf{\mu}}_{S_i} = -g \mu \hat{\mathbf{S}}_i$, where $\mu$ is a constant which can effectively count for different effects and $g$ is the $g$-factor (for example $g\approx2$ for the electron). 
We assume that the constant homogeneous magnetic field is applied along the $z$-direction
and the quantized electromagnetic mode 
along $z$-axis interacts with a number $p$ of spin magnetic moments only.
We can think of the spin chain as divided into three sections, 
with the middle section, consisting of $p$ spins, placed in the cavity.
Generally speaking, we expect the results to depend on the boundary conditions on the borderlines between the regions in and off the cavity.  Here we will suppose that the cavity won't affect directly the off spins. 
Moreover, we suppose that the spin chain resides in the $(x,y)$ plane,
and therefore, it is reasonable to work in the dipole approximation 
($\e^{\pm \i|\text{k}| \, \text{z}} \approx 1$).
The physical meaning of the latter is that the extent of the system in $z$-direction 
is much smaller than the wavelength of the mode, so the system experiences the same 
instantaneous field $B_{\pm}(z,t) \approx B_{\pm}(t)$ at any point. 
For convenience we depict schematically
our system in Fig. \ref{Fig:SystemGeometry}.

With the above assumptions we can write down the interaction term as
\begin{equation}
\hat{V}
	= \hbar \omega_1 \, \sum_{i} \hat{S}_{iz}
+ \i \hbar\Omega \sum_{i = 1}^p \left[\hat{S}_{iy} \left(\hat{a}_x - \hat{a}_x^{\dagger}\right) 
- \hat{S}_{ix} \left(\hat{a}_y - \hat{a}_y^{\dagger}\right) \right] .  \label{V_Cartesian}
\end{equation}
Here $\hat{a}_{x,y}$ and $\hat{a}_{x,y}^{\dagger}$ are the anihilation and creation 
operators for the linearly polarized field, $\Omega$ parametrizes the
coupling between the spins and the circularly polarized magnetic field,
and $\omega_1$ parameterizes the coupling between the spins and the
constant component of the magnetic field.
Note that if the sum in the first term runs over $1  \leq i \leq
p$, then $\hat{V}$ would coincide with the Dicke model \cite{Dicke1954}.

Because of the specificity of the problem it will be more convenient to introduce
a set of creation and annihilation operators for circularly polarized field defined via
\begin{equation}\label{polar}
\hat{a}_{\pm} = \mp\frac{1}{\sqrt{2}} (\hat{a}_x\mp \i\hat{a}_y) ,
\end{equation}
with the following canonical commutators
\begin{align}
[\hat{a}_{\eta},\hat{a}_{\zeta}] &= [\hat{a}_{\eta}^{\dagger},\hat{a}_{\zeta}^{\dagger}] = 0, \quad 
[\hat{a}_{\eta},\hat{a}^{\dagger}_{\zeta}] = \delta_{\eta\zeta} , \quad \eta,\zeta = +,-,
\end{align}
and the spin ladder operators 
\begin{equation}\label{spinladder}
\hat{S}_{i\pm} = \hat{S}_{ix} \pm \i\hat{S}_{iy} , \quad \hat{S}_{iz} = \hat{S}_{iz} , 
\end{equation}
that satisfy the commutation relations
\begin{equation}
[\hat{S}_{i+},\hat{S}_{j-}] = 2\delta_{ij}\hat{S}_{iz}, \, \qquad
[\hat{S}_{iz},\hat{S}_{j\pm}] = \pm\delta_{ij}\hat{S}_{i\pm} .
\end{equation}

In terms of the operators \eqref{polar} and \eqref{spinladder},
Hamiltonian \eqref{H_general} takes the form
\begin{subequations}\label{Hamiltonian}
\begin{align}
	\hat{H} = \ & \hbar\omega\left(\hat{n}_+ + \hat{n}_-  +
	\frac{p}{2} + \sum_{i=1}^p \hat{S}_{iz}\right) \notag \\
	& \ +J \sum_{i > j} \left(\hat{S}_{i+} \hat{S}_{j-} + \hat{S}_{i-} \hat{S}_{j+}\right) + \hat{H}^D,
	\label{Hamiltonian_a}
\end{align}
with
\begin{align}
	\hat{H}^D = \ & \hbar \omega_1 \,\sum_{i>p} \hat{S}_{iz} 
+ \hbar \Delta \,\sum_{i=1}^p \hat{S}_{iz} \nonumber \\
	&- \frac{\hbar\Omega}{\sqrt{2}} \sum_{i=1}^p \left[ (\hat{a}_+ + \hat{a}_-^{\dagger})
\hat{S}_{i+}  +  (\hat{a}_- + \hat{a}_+^{\dagger})
\hat{S}_{i-} \right], \label{H_D}
\end{align}
\end{subequations}
where we define the detuning $\Delta = \omega_1 - \omega$
and the number operators $\hat{n}_{\pm}=\hat{a}^{\dagger}_{\pm}\hat{a}_{\pm}^{}$. 
In Eq. \eqref{Hamiltonian}, and from now on, we explicitly
consider only this component of the chain with sites $i> 1$. The terms
involving sites with $i\le 0$ can be treated analogously. 
Moreover, note that we add $\tfrac p2\hbar\omega$ to the Hamiltonian in order to 
obtain the correct number of excitations operator.

We conclude this Section with a short discussion of the 
selection rules for the photon-induced transitions in the chain. 
First, we have 
\begin{align*}
\left[\hat{S}^2, \, \hat{H}^{D} \right] = 0, \,\,\,\,
\hat{S}^2 = \frac{1}{2} \left( \hat{S}_+\hat{S}_- + \hat{S}_-\hat{S}_+ \right) + \hat{S}_z^2 , \notag \\
\hat{S}_\pm = \sum_{i=1}^p \hat{S}_{i\pm},\,\,\,\,
\hat{S}_z = \sum_{i=1}^p \hat{S}_{iz},
\end{align*}
and so, 
the photon-induced transitions can occur only between states with same quantum number $S$.
Second, we notice that the projection of the total spin of the composite 
photons-spin chain system on the quantization axis $z$ is conserved. 
This can be proved by checking the commutator
\begin{align}
\left[\hat{n}_+ - \hat{n}_- + \hat{S}_z, \, \hat{H} \right] = 0 .
\end{align}
Therefore, since the photon carries a unit spin, the absorption/emission of a photon by the 
chain is accompanied by a change of the projection of the total spin of the chain $S_z$ by unity. 
Summing up, the selection rules for the total spin of the chain read
\begin{align}
\Delta S_z = \pm 1, \,\,\, \Delta S = 0. \label{Selection_Rules}
\end{align}

\section{Rotating wave approximation}\label{Sec_RWA}

The essence of RWA consists in neglecting 
corrections that are inversely proportional to the frequency $\omega$
and its higher degrees in the solution of the Scr\"{o}dinger equation.
In problems, like Jaynes-Cummings model  \cite{Jaynes1963}, 
it is usually assumed that the field is tuned near resonance, 
so that the detuning $\Delta = \omega_1 - \omega$ is 
orders of magnitude smaller than $\omega_1 + \omega \sim 2\omega$, 
the interaction constant $\Omega$ is much smaller than the photon
frequency, i.e.
$\alpha\Omega\ll\omega, \: \alpha \approx \sqrt{n}$,
and the  process time is much longer than the period of the field $T \gg 2\pi/\omega$.
For its $p$-spin generalization -- the Tavis-Cummings model -- one of
the conditions needs to be adapted, namely
\begin{align}
p \, \frac{\alpha_-\Omega}{\omega} \ll 1, \quad &\frac{\alpha_+\Omega}{\omega} \ll 1 , 
\quad \alpha_{\pm} \approx \sqrt{n_{\pm}} . \label{RWA_CondSmallDModified} 
\end{align}  
The condition (\ref{RWA_CondSmallDModified}) is best fulfilled 
in the case of left circularly polarized field-mode that we
will be studying hereafter.
Then, the right circularly polarized photons may appear only due to
counter-rotating terms. 
For a system of the size of $p \sim 10$ we will
assume $\alpha_- \lesssim 1$. 
Provided (\ref{RWA_CondSmallDModified}) are fulfilled the RWA can also
be applied to the Dickie component \eqref{H_D} of Hamiltonian
(\ref{Hamiltonian_a}) with $J \neq 0$ . 

Furthermore, we find that the same conditions, $T \gg 2\pi/\omega$ and
(\ref{RWA_CondSmallDModified}), are sufficient to ensure the validity
of the RWA for the Dicke component of
(\ref{Hamiltonian_a}) in the large detuning regime ($\Delta \to
-\omega$) as well.

We would like to keep the $XY$ term in Hamiltonian
(\ref{Hamiltonian_a}) unaffected by RWA. 
To this end, we consider the $XY$ chain in the nearest-neighbor approximation,  placed into a cavity (to avoid the contribution of the field-boundary terms, see Appendix \ref{RWA}),
and we obtain an additional sufficient condition, namely 
\begin{subequations}
\begin{align}
&p \, \frac{JT}{\hbar} \ll 1 , \quad \text{for} \quad \Delta\to 0 ,
\label{RWA_XY_condSmallD} \\
&p \, \frac{J}{\hbar\omega} \lesssim 1,  \quad \text{for} \quad \Delta\to -\omega.
\end{align} \label{RWA_J}
\end{subequations} 
Condition (\ref{RWA_XY_condSmallD}) seems very restrictive, however,  
for vanishingly small values of $J$ it is, obviously, fulfilled; 
for intermediate values of $J$, $J/\hbar\sim\Omega \ll\omega$
relevant to this study (\ref{RWA_XY_condSmallD}) is not fulfilled yet
RWA will not alter significantly the $XY$ interaction term;
for large $J$ the $XY$ component of Hamiltonian
(\ref{Hamiltonian_a}) dominates the dynamic process and the Dicke component (\ref{H_D}) should be treated as a perturbation.
Here we will turn a special attention to the large detuning regime.
A sketch of the derivation of conditions
\eqref{RWA_CondSmallDModified} and \eqref{RWA_J} is outlined in
Appendix \ref{RWA}. Let us point out, that more rigorous calculations might lead to less restrictive requirements.

In the framework of RWA there are a couple of important consequences, 
that help us simplify Hamiltonian (\ref{Hamiltonian}): 
First, we neglect any term containing right-hand circularly polarized components 
of the field that do not conserve the energy, such as 
$\hat{a}_-^{\dagger}\hat{S}_{i+}$ and $\hat{a}_-\hat{S}_{i-}$, 
and we set $\hat{n}_-=0$.
Second, since the number of excitations operator $\hat{N}$ commutes with 
the ensuing approximated Hamiltonian, it is an invariant quantity (a constant
of motion). 
We also need to remark that the RWA do not alter the selection rules (\ref{Selection_Rules}).

With the above simplifications, apart from the constant of motion $\hbar\omega\hat{N}$, the Hamiltonian of the $XY$ chain interacting  
with the photon mode splits into the sum of two terms: 
The Tavis-Cummings Hamiltonian $\hat{H}^{TC}$ \cite{Tavis1968} and the $XY$ spin Hamiltonian $\hat{H}^{XY}_{<p}$ \cite{Lieb1961a,E.Lieb1966b,DePasquale2008}, that are well known exactly solvable models. 
That is
\begin{subequations}\label{Hamiltonian_RWAt}
\begin{align}
\hat{H} &= \hbar\omega\hat{N} + \hat{H}^{XY}_{<p} + \hat{H}^{TC}, \label{Hamiltonian_RWA}
\end{align}
where
\begin{align}
\hat{N} &= \left(\hat{n}_+ 
+ \frac{p}{2} + \sum_{i=1}^p \hat{S}_{iz}\right) , \label{InvariantN} \\
\hat{H}^{XY}_{<p} &= J \sum_{p \geq i > j} \left(\hat{S}_{i+} \hat{S}_{j-} + \hat{S}_{i-} \hat{S}_{j+}\right),  \label{H_XY}\\
\hat{H}^{TC} &= \hbar \Delta \,\sum_{i=1}^p \hat{S}_{iz} 
- \frac{\hbar\Omega}{\sqrt{2}} \sum_{i=1}^p \left(\hat{a}_+ 
\hat{S}_{i+}  +  \hat{a}_+^{\dagger}\hat{S}_{i-} \right) .
\end{align}
\end{subequations}
Henceforth we omit the ``$+$'' subscript on the photon creation and annihilation operators
and will not be worried too much about the fulfillment of the
conditions for the validity of RWA.
The main goal will be to make the dependencies on the parameters more
lucid.

\begin{figure*}[ht!]
	\centering              
	\includegraphics[width=.9\textwidth]{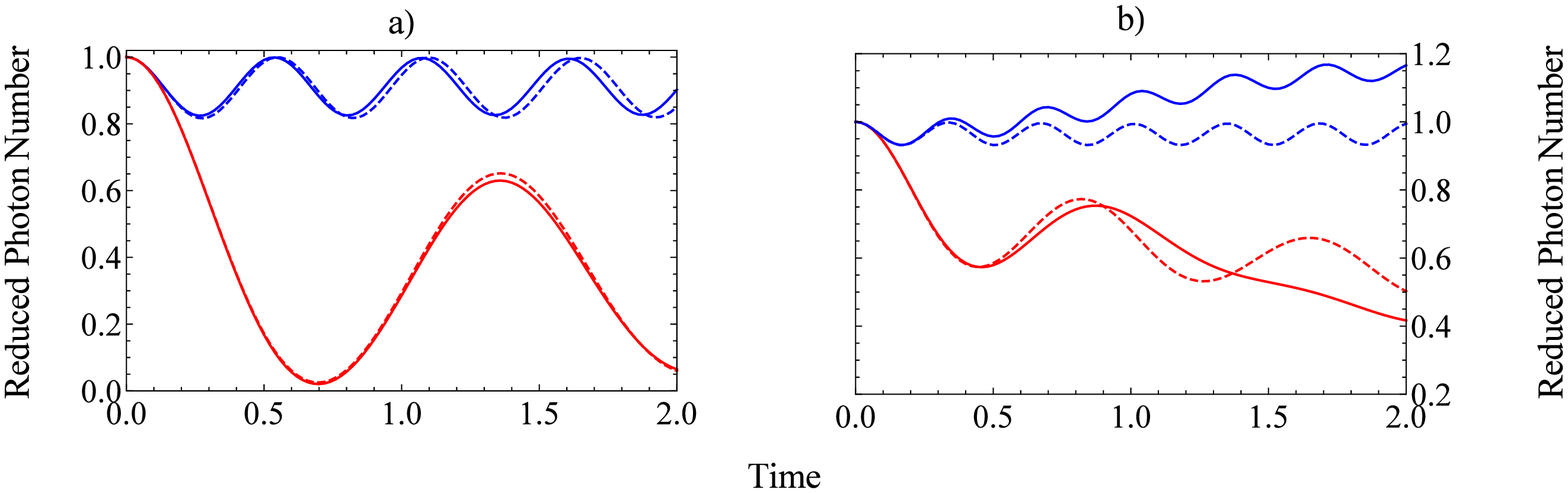}
	\caption{(Color online)  
		Reduced photon number $P_{\text{ph}}(t)$ as a function of dimensionless time $t/\tau$ ($\tau = 1/\Omega$) for various values of the control parameters is plotted in the figures. 
		$P_{\text{ph}}(t/\tau)$ obtained through the full Hamiltonian (\ref{Hamiltonian}) and the RWA-Hamiltonian (\ref{Hamiltonian_RWAt}) are given with solid and dashed curves respectively.  
		Parameters used for the computations are $p=12, \, N=4, \, J\tau/\hbar = 1, \, \omega\tau = 13, \Delta\tau = -12,-1 $ for (a) and $J\tau/\hbar = 3, \, \omega\tau = 1, \Delta\tau = 12,-0.1 $ for (b). 
		The curves corresponding to larger (smaller) by absolute values of the detununing are colored with blue (red). 
		In the full Hamiltonian diagonalization the state space of the left/right circularly polarized photons is truncated to six/three photons.  
	}
	\label{Fig:CheckRWA}
\end{figure*}

Examples illustrating the influence of the RWA on the reduced photon number $P_{\text{ph}}(t)$ are shown in  Fig. \ref{Fig:CheckRWA}. 
The curves in Fig. \ref{Fig:CheckRWA} a) represent $P_{\text{ph}}(t)$
when the experimental parameters fulfill (though not
strictly) the RWA conditions
\eqref{RWA_CondSmallDModified} and
\eqref{RWA_J}, while
those in Fig. \ref{Fig:CheckRWA} b) represent
$P_{\text{ph}}(t)$ when \eqref{RWA_CondSmallDModified} and
\eqref{RWA_J} are heavily violated.
It should be pointed out that $P_{\text{ph}}(t)$ exceeds unity in Fig. \ref{Fig:CheckRWA} b) which is due to the generation of (real) left circularly polarized photons by the terms $\hat{a}_+^{\dagger}\hat{S}_{i-}$,
after the virtual excitations induced by $\hat{a}_-^{\dagger}\hat{S}_{i+}$. 



\section{Diagonalization of $\boldsymbol{XY}$ component of the Hamiltonian}\label{Sec_HinXYBasis}

To be more specific, here and below, we consider the periodic $XY$
model in the nearest-neighbor approximation, and we assume an even 
number of sites $p$.
Note, that the periodic boundary conditions, $\hat{S}_{p+1\pm} =
\hat{S}_{1\pm}$, imply that the upper bound of the double sum in the
$XY$ Hamiltonian (\ref{H_XY}) must be changed from $i = p, \, j = p -
1$ to $i = p + 1, \, j = p$.

In order to analyze the Schr\"{o}dinger equation with Hamiltonian
(\ref{Hamiltonian_RWAt}), we transform $\hat{H}$
so that $\hat{H}^{XY}$ may be diagonalized
\cite{DePasquale2008, Tokihiro1993, Wu2018}.
The diagonalization procedure is outlined in Appendix
\ref{XY_Diagonalization_1}. This yiels
\begin{subequations}\label{HamiltonianRWA}
\begin{equation}
\hat{H} = \hbar\omega\hat{N} + \hat{H}(J,\Delta) - \frac{\hbar\Omega}{\sqrt{2}} \sum_{i=1}^p \left(\hat{a} \hat{S}_{i+} + \hat{a}^{\dagger}\hat{S}_{i-} \right),  \label{H_XY_diag}
\end{equation}
with
\begin{widetext}
\begin{align}
\hat{H}(J,\Delta) &= \sum_{\sigma = \pm} \frac{1+\sigma P}{2} \, \left\{ \sum_{k \in BZ} \left( 2J \Lambda_{k\sigma}  +   \hbar \Delta \right) \, \left(\hat{\eta}_{k\sigma}^{\dagger}\hat{\eta}_{k\sigma} - \frac{1}{2}\right) \right\} \frac{1+\sigma P}{2}  \qquad \text{and} \\
\sum_{i=1}^p  \hat{S}_{i+} &= \sum_{\sigma = \pm} \frac{1+\sigma P}{2} \, \sum_{k \in BZ} \hat{\eta}_{k\sigma}^{\dagger} \sum_{i=1}^{p} \phi^{(\sigma)\ast}_{ki} \exp \left(\i \, \pi \sum_{q,q^{\prime} \in BZ} \sum_{j = 1}^{j = i-1} \phi_{qj}^{(\sigma)\ast} \phi_{q^{\prime}j}^{(\sigma)} \hat{\eta}_{q\sigma}^{\dagger}\hat{\eta}_{q^{\prime}\sigma} \right) \frac{1-\sigma P}{2} ,  \label{SOME_eta}
\end{align}
\end{widetext}
\end{subequations}
where $\hat{\eta}_{k\sigma}$ and $\hat{\eta}^{\dagger}_{k\sigma}$ are Fermi operators obeying the canonical anti-commutation relations
\begin{align}
\{\hat{\eta}_{k\sigma}, \hat{\eta}_{k^{\prime}\sigma}^{\dagger}\} = \delta_{kk^{\prime}}, \quad
\{\hat{\eta}_{k\sigma}, \hat{\eta}_{k^{\prime}\sigma}\} = 
\{\hat{\eta}_{k\sigma}^{\dagger}, \hat{\eta}_{k^{\prime}\sigma}^{\dagger}\}  = 0,
\end{align}
$\phi^{\sigma}_{ki}$ are components of the orthonormal vectors $\phi^{\sigma}_{k}$,
\begin{align}
\quad \phi_{kj}^{\sigma} &= \frac{1}{\sqrt{p}}\,\e^{\i k^{\sigma}j}, \quad 
\sum_{i=1}^{p} \phi_{ki}^{\sigma} \phi_{k^{\prime}i}^{\sigma\ast} = \delta_{kk^{\prime}} , \label{Phi_orthogonallity}
\end{align}  
the energies (in units of $2J$) $\Lambda_{k\sigma}$ for the free $XY$ model read
\begin{align}
\Lambda_{k\sigma} &= \cos(k^{\sigma}) , 
\label{Lambda}
\end{align}
and the quasi-momentum summations run over the $p$ discrete values in the Brillouin zone ($BZ$)
\begin{align}  
k_{\eta}^{\sigma} &= -\pi + \left(2\eta + \frac{\sigma-3}{2}\right) \frac{\pi}{p} , \quad \eta=1,2,\dots, p.
\footnotemark
\label{Momentum_Definition}
\end{align}
\setcounter{footnote}{0}
\footnotetext{Recall that this formula is correct only for even $p$. The general result for quasi-momentum is  $k_{\zeta}^{\sigma} = -\pi + \left(2\zeta + p + \frac{\sigma+1}{2}\right) \frac{\pi}{p} , \quad \zeta=1,2,\dots, p $.}
The lower index in quasi-momentum is omitted everywhere except in
cases of summation over $\eta$ in Sec. \ref{SubSecA_FormalSolInFermionBasis}.

\section{Fermion Basis and physical quantities}\label{Sec_NotationsAndDefinitions}

Here, we define the quantum basis states where we further write down the matrix form of
the Schr\"{o}dinger equation and define the physical quantities
relevant to the description of the dynamics of the system.

\subsection{Basis states}

We consider the regime of strong magnetic field $B_z$ \footnote{Because
of the relation $\Delta \propto B_z$ the assumption for strong $B_z$
field  implicitly implies a restriction on the magnitude of $B_+$
field. In fact, the magnitude of $\Omega$ and hence of $|B_+|$ since
$\Omega \propto |B_+|$, has to be large enough (for instance
$\Omega/\Delta \gtrsim 0.1$) to achieve transfer of reasonable amount
of probability. In the text we have assumed that both the RWA and the
above conditions, bounding the magnitude of $|B_+|$ from above and
from below respectively, are fulfilled.},
set by the condition $\hbar\omega_1 \geq 2J$, 
when the ground state of the $XY$ model interacting with $B_z$
is the vacuum [free of $\eta$--quasi-particles
(ferromagnetic)] state, 
see Appendix \ref{XY_Diagonalization_21}.
Assuming that initially the excitations of 
the system are contained in the photon field and 
that the $XY$ model, subject to $B_z$, is in its ground state, 
we define the initial state to be the ($p+1$)-dimensional product state
\begin{align}
\vert N; 0\rangle \equiv \vert N\rangle \otimes \vert 0\rangle ,
	\qquad
\vert 0\rangle \equiv \vert 0, 0 , \dots, 0\rangle ,
\label{InitialState_1F}
\end{align}
where $N$ and $0$'s stand for the photon and $\eta_k$'s occupation
numbers, respectively.

Henceforth, we will also use the following shorthand notation 
\footnote{Below we use the notations of \textit{Wu} \cite{Wu2018}.}
\begin{subequations}
\begin{align}
\vert \overrightarrow{\eta}_m \rangle &= \prod_{l=1}^{m} \hat{\eta}^{\dagger}_{\eta_l\sigma_m} \vert 0 \rangle = \hat{\eta}^{\dagger}_{\eta_1\sigma_m}\hat{\eta}^{\dagger}_{\eta_2\sigma_m}\hat{\eta}^{\dagger}_{\eta_3\sigma_m}...\hat{\eta}^{\dagger}_{\eta_m\sigma_m} \vert 0 \rangle, \\
\overrightarrow{\eta}_m &\equiv (\eta_1,\eta_2, \dots,	 \eta_m), \quad 1 \leq \eta_1 < \eta_2 < \dots < \eta_m \leq p
\end{align}
\end{subequations}
for the eigenstates of the $XY$ chain.

The number of excitations $N$ is a conserved quantity, determined by the initial state.
So the dynamics of the system is restricted to the $D_{p,N}$-dimensional, 

\begin{align}
D_{p,N} = \sum_{m=0}^{\text{min}\{N,p\}} C_p^m, \,\qquad C_p^m = \frac{p!}{m!(p-m)!},
\end{align}
subspace of the Hilbert space ($N$-sector) spanned by the basis states 
\begin{align}
\{\vert N - m ;  \overrightarrow{\eta}_m \rangle \} .
\end{align}
Obviously, here $m$ stands for the number of excitations in the $XY$-chain.

An arbitrary state in the $N$-sector can be always given by the expansion
\begin{align}
\vert \psi_N (t) \rangle = \sum_{m=0}^{\text{min}\{N,p\}} \sum_{\overrightarrow{\eta}_m}
A_{\overrightarrow{\eta}_m}^{m}(t) \vert N - m ; \overrightarrow{\eta}_m \rangle ,
\end{align}
where $A_{\overrightarrow{\eta}_m}^{m}(t)$ are time-dependent probability amplitudes.

\subsection{Physical quantities}

We are interested in the influence of the spin chain on the
electromagnetic field and vice versa. To this end
we will compute the expectation values of the operators
corresponding to relevant physical quantities.
Therefore, to describe the evolution of the system in the
state $\vert \psi_N(t) \rangle$,
we consider the reduced photon occupation number \cite{Wu2018,Strater2012}
\begin{align}
P_{\text{ph}}(t) 
&= \frac{\langle \psi_N(t) \vert \hat{n} \vert \psi_N(t) \rangle}{N} = \, \sum_{m=0}^N \sum_{\overrightarrow{\eta}_m} \left(1 - \frac{m}{N}\right) \, P_{\overrightarrow{\eta}_m}^m(t), \\
P_{\overrightarrow{\eta}_m}^m(t) &= \vert A_{\overrightarrow{\eta}_m}^{m}(t) \vert^2 ,
\end{align}
and in addition, the (dimensionless) magnetization per site of the $XY$ spin chain 
\begin{align}
m(t) &= \frac{M(t)}{\mu p} =  - \frac{2}{p} \,\, \langle \psi_N(t) \vert \sum_{i = 1}^p \hat{S}_{iz} \vert \psi_N(t) \rangle = 1-\frac{2NP_\text{m}(t)}{p} , 
\label{M_PerSite} \\
P_{\text{m}}(t) &= 1 - P_{\text{ph}}(t) .
\end{align}
Note that Eq.~(\ref{M_PerSite}) is the result of the invariance of
the number of excitations operator Eq.~(\ref{InvariantN}), which also implies that $P_{\text{ph}}(t)$, $m(t)$, and $p/N$ are not independent quantities. 
The dependencies of $P_{\text{ph}}(t)$ and $m(t)$ on $p/N$, $p$ and $N$ 
in the regime of large detuning will be explored
in Subsec. \ref{SubSec_NumRes}.

\section{Single excitation} \label{Sec_SingleExcitation}

The case of a single excitation is of peculiar interest since its 
solution can be obtained in a closed form.
Moreover, this result will set the basis for our further considerations.  

As we start from the initial state (\ref{InitialState_1F}) with $N=1$, the dynamics of the system is restricted to the states describing single excitations. 
The only possibly non-vanishing transition matrix elements of
Hamiltonian \eqref{HamiltonianRWA} are 
\begin{align}
\langle 1_k ; 0 \vert \hat{H} \vert 1; 0 \rangle \equiv & \langle 0,
	\dots, 1_k, \dots, 0 ; 0 \vert \hat{H} \vert 1; 0, \dots, 0_k,
	\dots, 0 \rangle, \notag \\ 
	&\qquad \forall k \in BZ, \label{Transition_Matrix_Elements}
\end{align}
and their Hermitian conjugates. 
The exponents in (\ref{SOME_eta}) are effectively unity for these transitions. 
Thus they can be replaced by mathematically identical (effectively unit) 
Jordan-Wigner string operators 
\footnote{We silently set $\sigma=-1$ between the projectors.},
\begin{equation}
\exp \left(\pm \i \, \pi \sum_{q,q^{\prime} \in BZ} \sum_{j = 1}^{i-1} \phi_{qj}^{\ast} \phi_{q^{\prime}j} \hat{\eta}_{q}^{\dagger}\hat{\eta}_{q^{\prime}}  \right) \rightarrow \,\,
\exp\left(\pm \i\pi \sum_{q = -\pi}^{k-2\pi/p} \hat{\eta}_{q}^{\dagger}\hat{\eta}_{q} \right) . 
\end{equation}
Then by means of Jordan-Wigner transformations we define a set of half-spin operators
\begin{subequations}
\begin{align}
\hat{\Sigma}_{k-} &= \exp\left( -\i\pi \sum_{q = -\pi}^{q = k-2\pi/p}
	\hat{\eta}_{q}^{\dagger}\hat{\eta}_{q}  \right)
	\hat{\eta}_{k}, \\
	\hat{\Sigma}_{k+} &= \hat{\eta}_{k}^{\dagger} \exp\left( \i\pi \sum_{q = -\pi}^{q = k-2\pi/p} \hat{\eta}_{q}^{\dagger}\hat{\eta}_{q} \right), \\
\hat{\Sigma}_{kz} &= \hat{\eta}_{k}^{\dagger} \hat{\eta}_{k} - \frac{1}{2},
\end{align}
\end{subequations}
obeying the following commutation relations
\begin{align}
[\hat{\Sigma}_{q+},\hat{\Sigma}_{q^{\prime}-}] = 2\delta_{qq^{\prime}}\hat{\Sigma}_{qz}, \quad 
[\hat{\Sigma}_{qz},\hat{\Sigma}_{q^{\prime}\pm}] = \pm\delta_{qq^{\prime}}\hat{\Sigma}_{q\pm} .
\end{align}

With the aid of these operator variables we can rewrite Hamiltonian
(\ref{HamiltonianRWA})
in the following form
\begin{subequations}
\begin{equation}
\hat{H} =\hbar\omega\hat{N} + \sum_{k \in BZ, \, k\neq 0} P_k
\hat{\Sigma}_{kz} +\hat{H}_{JC} , \label{H_to_JC}
\end{equation}
with
\begin{equation}
\hat{H}_{JC} = \,\, P_0 \hat{\Sigma}_{0z}  + \, \frac{Q}{2} \left(\hat{a}  \hat{\Sigma}_{0+} + \hat{a}^{\dagger}  \hat{\Sigma}_{0-} \right) ,  \label{H_JC}
\end{equation}
and
\begin{equation}
P_k = 2J\Lambda_k + \hbar \Delta, \qquad Q = - \hbar\Omega \sqrt{2p} .
\label{Hamiltonian_To_JC}
\end{equation}
\end{subequations}

The second term in \eqref{H_to_JC} is also a constant of motion
\begin{align}
\left[\sum_{k \in BZ, \, k\neq 0} P_k \hat{\Sigma}_{kz} , \hat{H}\right] = 0 ,
\end{align}
that describes the free evolution of two-level systems with excitation
energies $P_k$.
The third term, i.e. $\hat{H}_{JC}$, given explicitly in \eqref{H_JC},
describes the field-induced dynamics of the $XY$ system. 
It corresponds to the Jaynes-Cummings model for a two-level system 
interacting with a single-mode quantized cavity field.

Since only the zero-momentum mode is involved in the interaction described by (\ref{H_JC}), 
the dynamics is confined to the two-dimensional subspace
spanned by
\begin{subequations}
\begin{equation}
\{ \vert \Phi_\text{in}\rangle , \vert 0; \tfrac{1}{2}_{\small 0} \rangle \} ,
\end{equation}
with 
\begin{align}
&\vert \Phi_\text{in}\rangle \equiv \vert 1; -\tfrac12\rangle
	\equiv\vert 1; -\tfrac12, -\tfrac12 , \dots, -\tfrac12\rangle , 
\label{InitialState_1F_Sigma} \\
&\vert 0; \tfrac{1}{2}_{{\small 0}}\rangle \equiv \vert 0; -\tfrac12, -\tfrac12 ,
\dots, \tfrac{1}{2}_{k=0}, \dots, -\tfrac12\rangle  ,
\end{align}
\end{subequations}
where the spin chain subsystem is contained in the $\Sigma$--representation.

The solution of the Schr\"{o}dinger equation with Hamiltonian $\hat{H}_{JC}$ (\ref{H_JC})
and initial condition $\vert \Phi_\text{in}\rangle$ (\ref{InitialState_1F_Sigma}) reads
\cite{Jaynes1963}
\begin{align}
\vert \Phi(t) \rangle 
= \ & -\i \sin\left(\frac{\ce_+ t}{\hbar}\right) \sin
	\left(\phi\right) \vert 0; \tfrac{1}{2_0}  \rangle \nonumber \\
& + 
\left[ \i \sin\left(\frac{\ce_+ t}{\hbar}\right) 
\cos\left(\phi\right) + \cos\left(\frac{\ce_+ t}{\hbar}\right) \right]
	\vert 1; -\tfrac{1}{2}\rangle,
\label{Solution_JC} 
\end{align}
where the angle $\phi$ and the eigenvalues $\ce_{\pm}$ are given by
\begin{subequations}\label{ham_sol}
\begin{align}
\phi &= \tan^{-1} \left(\frac{Q}{P_0}\right) \label{phi_angle} , \\
\ce_{\pm} &= \pm\frac{1}{2}\sqrt{P_0^2 + Q^2} . \label{E_pm}
\end{align}
\end{subequations}

We can now write down the solution
of the Schr\"{o}dinger equation with Hamiltonian
(\ref{H_to_JC}) and initial condition
\begin{align}
\vert\psi_\text{in}\rangle & \equiv\vert 1; -\tfrac12, -\tfrac12 ,
	\dots, -\tfrac12\rangle .
\end{align}
This is obtained by recovering the constants that add up to the phase 
\begin{align}
\vert \psi\rangle &= \exp\left(- \frac{\i t}{\hbar} \, \hat{H} \right)  \vert \psi_\text{in}\rangle \notag \\ 
	&= \exp \left\{ - \frac{\i t}{\hbar} \left[ \frac{P_0}2-\sum_{k\in BZ}
	\frac{P_k}{2} +
	\hbar\omega \left( 1 - \frac{p}{2} \right) \right] \right\} \vert \Phi(t)\rangle . \label{psi_0_order}
\end{align} 
Note that the initial condition $\vert\psi_\text{in}\rangle$ takes the
same form as $\vert\Phi_\text{in} \rangle$. 
To avoid any confusion, we remind that the former is related to a real-space 
configuration in the spin representation (where $\hat{S}_i$
act), 
while the latter is related to a momentum space configuration in the spin 
$\Sigma$--representation (where $\hat{\Sigma}_k$
act).
Since both reflect the same physical state, though in different bases,
we have $\vert\psi_\text{in} \rangle \equiv \vert\Phi_\text{in} \rangle$.

From (\ref{Solution_JC}) we have for the time-dependent probabilities
of the excitation in the zero-momentum mode and in the field-boson,
respectively,
\begin{align} 
 P_{\text{m}0}(t) &= \langle \Phi(t) \vert \left( \hat{\Sigma}_{0z} +
	\frac{1}{2} \right) \vert \Phi(t) \rangle \label{Pm_SinglePhoton} \notag \\ 
	&= \frac{Q^2}{P_0^2+Q^2} \,  \sin^2\left(\frac{\ce_+t}{\hbar}\right)  , \\
 P_{\text{ph}}(t) &=  1 - P_{\text{m}0}(t).
\end{align}
The probabilities oscillate, reflecting the transfer of excitation between the
field and the spin chain, with frequency $\ce_+/\hbar$, and an
amplitude determined by the angle $\phi$.
It is clear, see definitions (\ref{ham_sol}), that with an increase of the
absolute value of the detuning $|P_0|$ the amplitude of the
oscillation decreases while its frequency increases. 
Obviously, the magnetization per site (\ref{M_PerSite}) also reveals oscillatory 
behavior defined by $P_{\text{m}0}(t)$.

A remark concerning the range of validity of the results 
in the large detuning regime is in order.
The amplitude of the oscillation depends on the ratio $Q/P_0 \sim \sqrt{2p}\Omega/\Delta $, 
combined with the RWA condition $p\Omega/\Delta \ll 1$ 
leads to the requirement $Q/P_0 \ll \sqrt{2/p}$. 
The latter imposes an upper bound on $Q/P_0$ for which our considerations are valid.
The bound increases as the number of spins $p$
decreases and thus the amplitude of the oscillations of
$P_{\text{ph}}(t)$ and the magnetization $m(t)$ might become larger within RWA.

\begin{figure}[t]
	\pix{width=.9 \columnwidth}{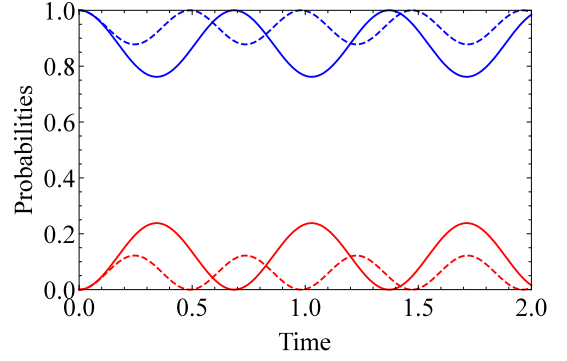}
	\caption{(Color online) 
		Dynamics of the probabilities for an excitation in the field, $P_\text{ph}(t/\tau)$ ($\tau = 1/\Omega$)  blue curves, and in the zero-momentum mode, $P_\text{m}(t/\tau)$, red curves, for a chain of $p=10$ spins and $\,N=1$ excitation. 
		The coupling $J$ and the detuning 
		are $J\tau/\hbar=1$ and $\Delta\tau=-10$, respectively. 
		The solid curves correspond to the anti-ferromagnetic case ($J>0$), and the dashed curves to the ferromagnetic case ($J<0$).
		The dimensionless time $t/\tau$ is
		shown on the abscissa.}
	\label{Fig:SingleExcitation}
\end{figure}

An example of dynamics of the probabilities for an excitation is
depicted in Fig. \ref{Fig:SingleExcitation}. 
Two cases are considered -- anti-ferromagnetic ($J>0$) and ferromagnetic ($J<0$).
Due to the larger absolute value of the detuning $|P_0|$, the
amplitude is smaller and the frequency is higher in the case of a ferromagnetic spin chain.
The physical reason to have larger $|P_0|$ is that the energy $-2|J|\Lambda_0$ of the 
zero-momentum mode in the ferromagnetic $XY$ chain is lower.   
Even for the anti-ferromagnetic chain, the reduction of the magnetization does not exceed $5\%$, for this specific choice of parameters.

\section{General (multiple excitation) case} \label{Sec_GeneralCase}

Here we recall the explicit form of the Hamiltonian matrix elements
and the formal solution of the problem,
which are then used for numerical computations.

\subsection{Formal solution} \label{SubSecA_FormalSolInFermionBasis}

In the general case of multiple
excitations, when $N \leq p$, the matrix elements of
Hamiltonian (\ref{HamiltonianRWA}), up to an irrelevant constant, are given by
\footnote{The formula for the transition matrix elements is first
given in \cite{Tokihiro1993}. Another derivation is published later
in \cite{Wu2018}.}
\begin{align} \label{MatrixElement_FermionBasis}
\langle \overrightarrow{\eta}_{m^{\prime}}^{\prime} ; N - m^{\prime}
	\vert & \hat{H} \vert  N - m  ; \overrightarrow{\eta}_m \rangle \notag \\
	& \quad = \ \delta_{mm^{\prime}}\prod_{l=1}^m\delta_{\eta_l^{\prime}\eta_l}[\mathcal{E}_{\overrightarrow{\eta}_m} + \hbar \omega (N - m)] \notag \\
	&\qquad - \delta_{m^{\prime}m+1} \, \frac{\hbar\Omega}{\sqrt{2}} \, F^{\ast}_{\overrightarrow{\eta}^{\prime}_{m+1}\overrightarrow{\eta}_m} \sqrt{N - m} \notag\\
&\qquad - \delta_{m^{\prime}m-1} \, \frac{\hbar\Omega}{\sqrt{2}} \,
	F_{\overrightarrow{\eta}_{m}\overrightarrow{\eta}^{\prime}_{m-1}}
	\sqrt{N - m + 1}.
\end{align}
With
\begin{align}
\mathcal{E}_{\overrightarrow{\eta}_m} = \sum_{l=1}^m \varepsilon_{\eta_l} , \quad \varepsilon_{\eta} &= \hbar\omega_1 + 2J\Lambda_{k_{\eta}} ,
\end{align}
the total energy of the $m$
excitations of the $XY$ chain in the state $\vert
\overrightarrow{\eta}_m \rangle$.
$F_{\overrightarrow{\eta}_{n+1}\overrightarrow{\eta}^{\prime}_{n}}$ is
the collective spin operator matrix
element
\begin{align}
F_{\overrightarrow{\eta}_{n+1}\overrightarrow{\eta}^{\prime}_{n}}
	&\equiv  \langle  \overrightarrow{\eta}_{n}^{\prime} \vert
	\sum_{j=1}^{p} \hat{S}_{j-} \vert \overrightarrow{\eta}_{n+1}
	\rangle \notag \\
	&=  2^n p^{\frac{1}{2}-n} \delta\left(\Delta_{\overrightarrow{\eta}_{n+1},\overrightarrow{\eta}^{\prime}_n},0 \right) h_{\overrightarrow{\eta}_{n+1},\overrightarrow{\eta}^{\prime}_n} , 
\end{align}
where
\begin{align}
\delta(x,y) =\,\, \left\{
\begin{tabular}{l}
$1 \quad \text{if} \,\,   x - y = 2\pi r, \,\,\,  r \in \mathbb{Z}$, \\
$0 \quad \text{otherwise} , $
\end{tabular}
\right.
\end{align}
is the Kronecker delta function,
\begin{align}
\Delta_{\overrightarrow{\eta}_{n+1},\overrightarrow{\eta}^{\prime}_n} &= \sum_{j=1}^{n+1} k_{\eta_j}^{\sigma_{n+1}} - \sum_{i=1}^{n} k_{\eta_i^{\prime}}^{\sigma_n} 
\end{align}
is the momentum transfer between the corresponding
states, and 
\begin{align}
h_{\overrightarrow{\eta}_{n+1},\overrightarrow{\eta}^{\prime}_n} &= \frac{\prod_{i>i^{\prime}} \left(\e^{-\i k_{\eta^{\prime}_i}^{\sigma_n}} - \e^{-\i k_{\eta^{\prime}_{i^{\prime}}}^{\sigma_n}} \right) \prod_{j>j^{\prime}} \left( \e^{\i k_{\eta_j}^{\sigma_{n+1}}} - \e^{\i k_{\eta_{j^{\prime}}}^{\sigma_{n+1}}} \right)}{ \prod_{i=1}^{n}\prod_{j=1}^{n+1} \left[1-\e^{-\i\left(k_{\eta_j}^{\sigma_{n+1}} - k_{\eta_i^{\prime}}^{\sigma_{n}}\right)} \right]} .
\end{align}
Note, that for non-vanishing off-diagonal matrix elements 
the number of $\eta$-excitations can only differ by unity.

With this at hand, for any $N$-sector, we can write the
Schr\"{o}dinger equation in matrix form in the basis $\{\vert N - m ;
\overrightarrow{\eta}_m \rangle \}$, and integrate it to find the
probability amplitudes $A(t)$. Formally, we have
\begin{align}
A(t) = \exp\left(-\frac{\i t}{\hbar} H\right) A_{\text{in}} , \quad A_{\text{in}} = \left(1,0,0, \dots, 0 \right)^T . 
\end{align}

\subsection{Numerical results} \label{SubSec_NumRes}


\begin{figure*}
	\centering
	\includegraphics[width=0.9\textwidth]{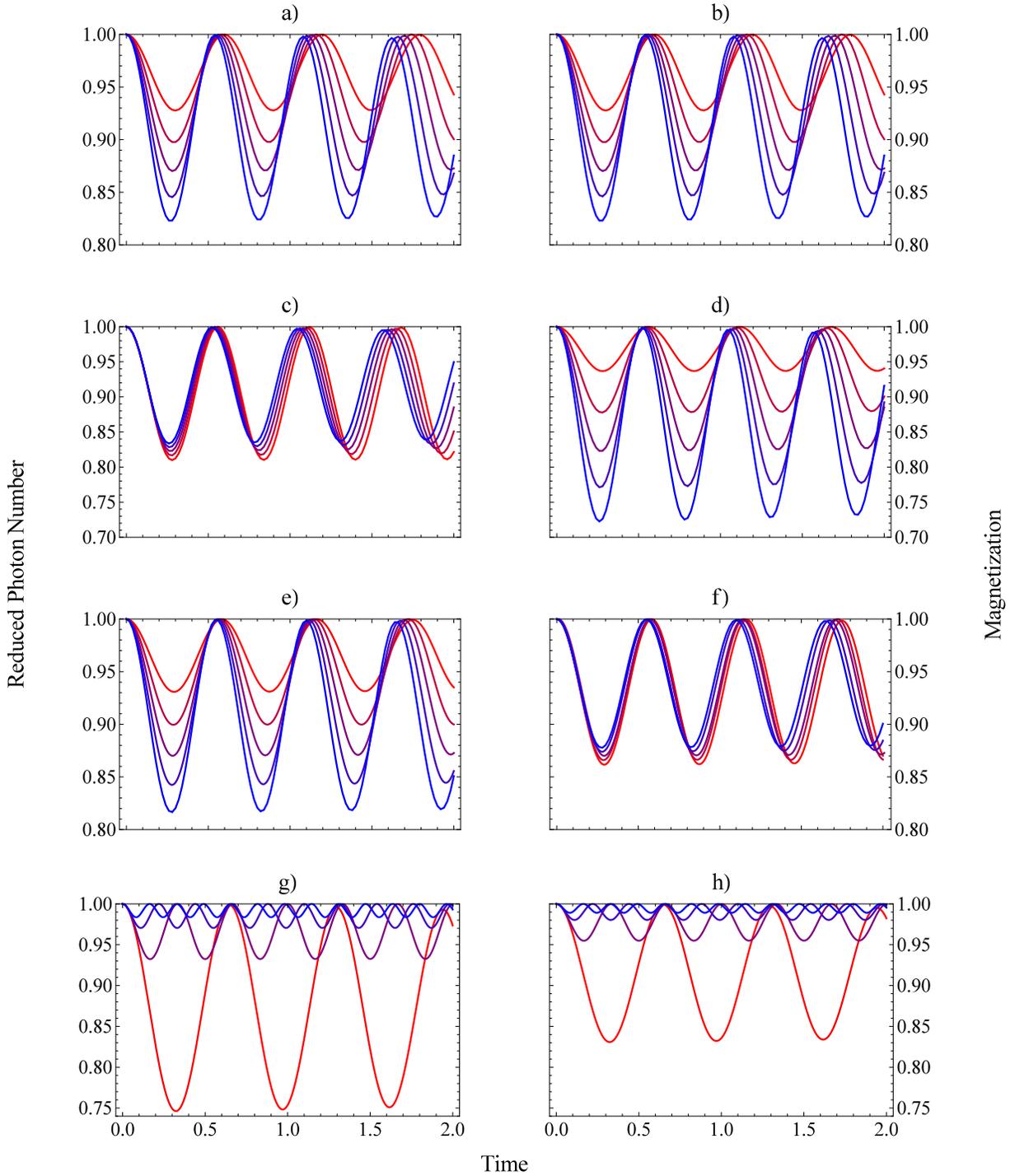}
	\caption{(Color online) 
		Dynamics of the reduced photon occupation number $P_\text{ph}(t/\tau)$ ($\tau = 1/\Omega$) in left panel and (dimensionless) magnetization per site $m(t/\tau)$ in right panel for various values of the control parameters.
		In (a) and (b), both the number of excitations and the number of the spins in the chain are varied simultaneously $(N,p) = (2,4),(3,6), \dots, (6,12)$, keeping the ratio $p/N =2$ fixed. 
		The values of the other parameters are $\,J\tau/\hbar=1,\, 
		\, \Delta\tau = -12$. 
		In (c) and (d), the number of excitations is varied in the range $N = 2,4, \dots, 10$.
		We choose $p = 12$ and the values of the parameters are the same as for the ones for (a) and (b).
		In (e) and (f), the number of spins in the chain takes the values $p=4,6, \dots, 12$.  
		We choose $N = 4$ and the values of the parameters are the same as for the ones for (a) and (b).
		In (g) and (h) the detuning varies in the range $\Delta\tau=-10,-20,-30,-40$ and the values of the other parameters are $N=4,\,p=12,\,J\tau/\hbar=1$.
		The increase of the absolute value of the components
		varied is encoded with a change of the color from red to blue.   
		The dimensionless time $t/\tau$ is plotted on the abscissas. }
	\label{Fig:MultipleleExcitations}
\end{figure*}


The dynamics of the reduced photon number $P_\text{ph}(t)$ and magnetization per site $m(t)$ 
for various values of the number of spins $p$, total number of excitations $N$, 
and detuning $\Delta$, is shown in Fig. \ref{Fig:MultipleleExcitations}. 
Particular cases near resonance $\Delta\tau \ll 1$ ($\tau = 1/|\Omega|$) 
are not considered here, while the on-resonance regimes will be discussed 
in Sec. \ref{Sec_PhotonEmission}. 
In this Section we explore the regime of large detuning $\Delta\tau \gg 1$.

As a general remark we see that Fig. \ref{Fig:MultipleleExcitations}
reveals an oscillatory behavior of $P_{\text{ph}}(t)$ and $m(t)$ with
frequency slightly dependent on the values of $p$ and $N$.
In fact, the frequency depends mostly on $\Delta$ and $J$, for fixed
coupling $\Omega$.

In Fig. \ref{Fig:MultipleleExcitations} a) several curves represent
the behavior of
$P_\text{ph}(t)$ for different values of $p$ and $N$,
at fixed ratio $p/N$. 
With the increase of their values the amplitude also increases, 
because the larger part of the excitations is exchanged between the
electromagnetic field and the chain. 
The curves in Fig. \ref{Fig:MultipleleExcitations} b) represent the magnetization $m(t)$ for the same 
set of values $p$ and $N$ as in Fig. \ref{Fig:MultipleleExcitations} a). 
The figure is completely analogous to Fig. \ref{Fig:MultipleleExcitations} a), that follows from (\ref{M_PerSite}), 
where for $p/N=2$ we obtain $P_\text{ph}(t)=m(t)$. 

The dynamics of $P_\text{ph}(t)$ for different values of $N$ and fixed $p$ is shown in Fig. \ref{Fig:MultipleleExcitations} c). 
We observe that the amplitude of $P_\text{ph}(t)$ is almost independent on $N$. 
In Fig. \ref{Fig:MultipleleExcitations} d) the dynamics of the respective magnetization $m(t)$ is shown. 
Contrary to the $P_\text{ph}(t)$, the amplitude of $m(t)$ is increasing 
against $N$. 
We see exactly the opposite behavior in Figs. \ref{Fig:MultipleleExcitations} e) and \ref{Fig:MultipleleExcitations} f)
where the dynamics of $P_\text{ph}(t)$ and $m(t)$ are shown for different values of $p$ and fixed $N$. 
This can be understood from expression (\ref{M_PerSite})
that shows a symmetry with respect to simultaneous exchange of $P_{\text{ph}}(t)$ and $m(t)$ together with $p$ and $2N$. 

In Figs. \ref{Fig:MultipleleExcitations} g) and \ref{Fig:MultipleleExcitations} h) we observe an oscillatory behavior where the frequency of both 
$P_{\text{ph}}(t)$ and $m(t)$ is increasing, while their amplitudes are decreasing, 
with the increase of the absolute value of the detuning.

\begin{figure}[ht!]
   \centering
\begin{tabular}{c}
   \includegraphics[width=.9 \columnwidth]{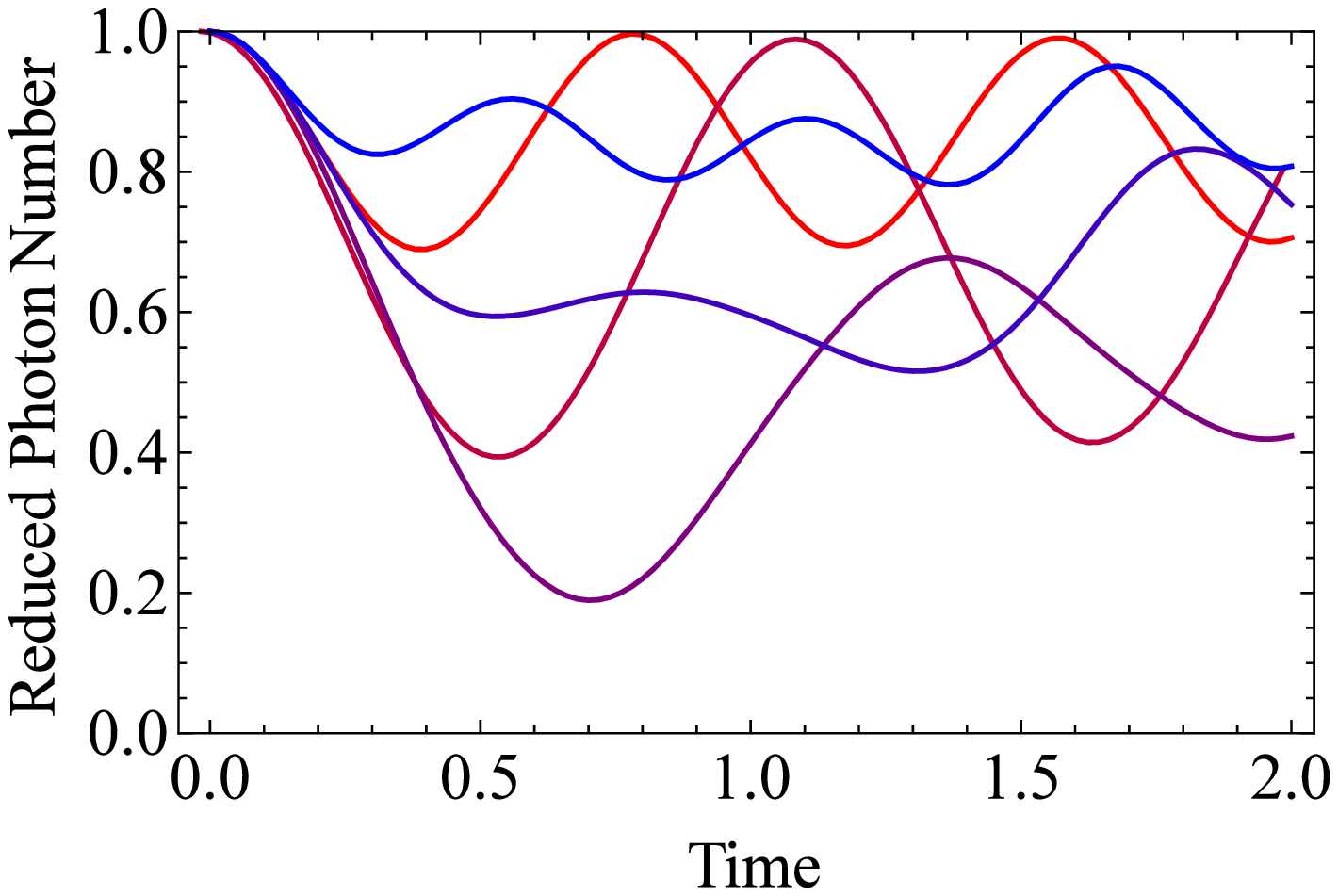} 
\end{tabular}
   \caption{(Color online)   
    Dynamics of the reduced photon occupation number $P_\text{ph}(t/\tau)$ ($\tau = 1/\Omega$) for $N=4,p=10$, $\Delta\tau = -10$, and various values of the spin-spin coupling $J\tau/\hbar = 2,4,\dots, 10$. 
    The dimensionless time $t/\tau$ is plotted on the abscissa. 
	The increase of $J$ is encoded with a change of the color from red to blue. }
   \label{Fig:MultipleleExcitationsJ}
\end{figure}

The dependence of $P_{\text{ph}}(t)$
against $J$ for relatively large values
is shown for completeness in Fig. \ref{Fig:MultipleleExcitationsJ}. 
Both the amplitude and frequency depend in a nontrivial manner on $J$. 
The behavior of the magnetization $m(t)$ (not shown in the figure) 
is qualitatively similar to that of $P_{\text{ph}}(t)$.

\section{Effective Hamiltonians}\label{Sec_EffectiveHamiltonians}

In Sec. \ref{Sec_SingleExcitation}, we showed
in the single-excitation case
that the excitation is \textit{shared} among the field and 
the zero-momentum mode of the chain. In the case of multiple
excitations, a close inspection of the transition matrix elements in
(\ref{MatrixElement_FermionBasis}), similarly, leads us to the
conjecture that the majority of excitations are shared between
the chain modes of zero net momentum and the field. Applying this
conjecture, we will be able to identify two somehow different
perturbative expansions. These are expressed in terms
of the expansions of
the perturbed tridiagonal Hamiltonian matrix model and of the perturbed
Tavis-Cummings model. Below, we will use only
unperturbed Hamiltonians,
and will regard them as reference effective Hamiltonians
reproducing approximately the short-time evolution in the case of
small number of total excitations. The reason not to deal with the
correction terms is that they cannot be expressed by simple analytic formulas.
Depending on the needed accuracy, at some point it may become
much more practical merely to run a full numerical simulation. 

We would like to anticipate that the selection rules (\ref{Selection_Rules}) 
for the derived effective Hamiltonians remain valid. 
Moreover, we verified numerically that
the named effective Hamiltonians 
are less accurate in the near resonant case.

\subsection{Effective tridiagonal  Hamiltonian} \label{TridiagonalHEff}

The simplest way to apply our conjecture is to restrict the states
involved in the dynamics, in zeroth order, to those that possess the
set of the smallest (in absolute value) possible momenta for the given
number of excitations in the chain. Now, we can build the zeroth-order
Hamiltonian $H_0$ as a sparse matrix with non-vanishing elements --
couplings and detunings, determined by the matrix elements between them. 
Explicit expressions for the latter can be found by means of
(\ref{MatrixElement_FermionBasis}).
To be specific, $H_0$ connects only the states $\vert
N;\overrightarrow{\eta}_0 \rangle \leftrightarrow \vert N-1;
\overrightarrow{\eta}_1 \rangle \leftrightarrow \vert N-2;
\overrightarrow{\eta}_2 \rangle \leftrightarrow \vert N-3;
\overrightarrow{\eta}_3 \rangle  \leftrightarrow \vert N-4;
\overrightarrow{\eta}_4 \rangle \leftrightarrow \dots$, where
\begin{align} \label{TridiagHEffHSpace}
\overrightarrow{\eta}_0 &= (\dots, 0, \dots) , \notag \\
\overrightarrow{\eta}_1 &= (\dots, 0,\tfrac p2+1,0, \dots) , \notag \\
\overrightarrow{\eta}_2 &= (\dots, 0,\tfrac p2,\tfrac p2+1,0, \dots) , \notag \\
\overrightarrow{\eta}_3 &= (\dots, 0,\tfrac p2,\tfrac p2+1,\tfrac p2+2,0, \dots) , \notag \\
\overrightarrow{\eta}_4 &= (\dots, 0,\tfrac p2-1,\tfrac p2,\tfrac
	p2+1,\tfrac p2+2,0, \dots) \,.
\end{align}
Thence the full Hamiltonian matrix is written in the form
\begin{align} \label{pert}
H = H_0 + V, 
\end{align}
and we can proceed (keeping the full Hilbert space) with the perturbation theory in a standard
manner. Alternatively, weakly populated states are ruled out
by means of adiabatic elimination
and the correction term $V$ accounts for an additional effective detuning.
Then, the order of the correction to the tridiagonal effective model
is determined roughly to be $ \langle TME \rangle^2/\Delta \sim
\Omega^2/\Delta$, where $\langle TME \rangle$ stands for the (order
of) expectation value of the transition matrix elements between the
states involved in the dynamics
and those eliminated states with
respect to the eigenvectors of $H_0$.
It is seen, that in general the eigenvalues and eigenstates of
$H_0$, and so the correction terms have to be calculated numerically.

The unperturbed Hamiltonian $H_0$ in \eqref{pert}, however, can be
considered as an effective Hamiltonian approximating the actual
problem, for a couple of
periods and a small total number of
excitations. After excluding the zero rows and columns, the effective
Hamiltonian is greatly reduced in size to a $(N+1) \times (N+1)$ matrix
and takes a tridiagonal form. As we have already seen in Sec.
\ref{Sec_SingleExcitation} for a single excitation, $H_0$ actually
generates the exact dynamics. With an increase of $N$ the accuracy of
this approximation decreases, because of the increasing number of
possible routes from which the probability can leak out of the
subspace spanned by the states $\vert N;\overrightarrow{\eta}_0
\rangle,
\vert N-1; \overrightarrow{\eta}_1 \rangle, 
\vert N-2; \overrightarrow{\eta}_2 \rangle, 
\vert N-3;  \overrightarrow{\eta}_3 \rangle, 
\vert N-4;  \overrightarrow{\eta}_4 \rangle,  \dots \,$ . 
In principle, with the increase of $N$, one can gradually add the most
involved in the dynamics modes and still find relatively simple, more
accurate, and compact expressions of the effective Hamiltonians.

\begin{figure}[th!]
	\pix{width=.9 \columnwidth}{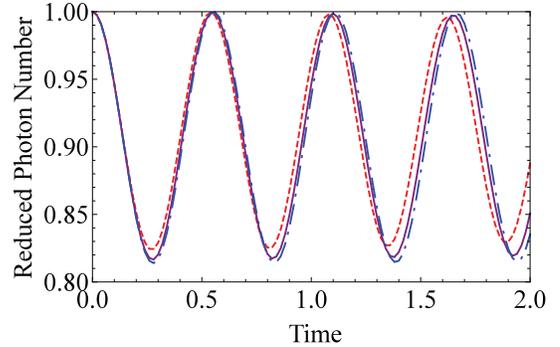}
	\caption{(Color online) Dynamics of the reduced photon number $P_{\text{ph}}(t/\tau)$ ($\tau = 1/\Omega$)  -- exact, represented with a solid purple curve; approximate, obtained with the use of the effective tridiagonal 
	(subsec. \ref{TridiagonalHEff}) and Tavis-Cummings (subsec.
		\ref{TCHEff}) Hamiltonians and represented with dashed
		red and dashed-dotted blue curves, respectively. 
	The parameters used for the computations are $N=4,\,p=12,\,J\tau/\hbar=1,\, 
	\Delta\tau=-12$.
	The dimensionless time $t/\tau$ is plotted on the abscissa.
	}
	\label{Fig:EffectiveHEvolution}
\end{figure}

For instance, in Fig. \ref{Fig:EffectiveHEvolution} we present both
the results obtained with the use of the effective tridiagonal model
and the exact solution. Comparison of the curves shows that for this
specific choice of parameters, and for short-time dynamics, the
accuracy of the approximated from the exact reduced photon number
is around $5\%$.

\subsection{Effective Tavis-Cummings model}\label{TCHEff}

Another possible choice for the effective Hamiltonian can be obtained
under the assumption of large enough number of sites $p$ when we can
neglect the periodic boundary term $\hat{H}_b$ (\ref{H_b})
being of
order $O(p^{-1})$. This is equivalent to work only in the $\sigma = -
1$ subspace. Further, since the states most involved in the dynamics
are those with zero net momentum of the chain, we can find the
zero-order Hamiltonian simply setting $k=0$, which is $\Lambda_k=1, \,
\forall k$. Switching back to the spin-$\hat{S}$ representation, we
see that this is nothing but the (exactly solvable)
Tavis-Cummings model
\begin{align}
\hat{H}_0 &= \,\, \hbar\omega\hat{N} +  (2J + \hbar \Delta) \hat{S}_{z} - \frac{\hbar\Omega}{\sqrt{2}} \left(\hat{a}_+ \hat{S}_+  +  \hat{a}_+^{\dagger} \hat{S}_- \right). \label{HEffTC} 
\end{align}
As we work in the large detuning limit, Hamiltonian
(\ref{HEffTC}) can easily be (approximately) diagonalized by
means of the method proposed in Ref. 
\cite{Klimov_2002}, but we are not going to pursue this route here.
The first non-vanishing correction to the Hamiltonian $\hat{H}_0$ with
respect to the small wave number $k$ is of second-order
$-J\,\sum_{k\in BZ} k^2 \langle\hat{\eta}^{\dagger}_k\hat{\eta}_k
\rangle$.
Recall that with the increase of $k$ the states are poorly populated,
which additionally suppresses the magnitude of the correction term.  
Here again, the correction terms to the effective Hamiltonian will be given either 
by complex analytical formulas or computed numerically. 
For that reason we will not deal with the corrections,
but rather will treat $\hat{H}_0$ as an effective approximation. The
effective Tavis-Cummings Hamiltonian approximates well (even better
than the tridiagonal one) the actual problem for small values of $N$
and for short time scales as seen in Fig. \ref{Fig:EffectiveHEvolution}.

\section{Photon emission}\label{Sec_PhotonEmission}

\begin{figure}[ht!]
	\centering              
	\includegraphics[width=\columnwidth]{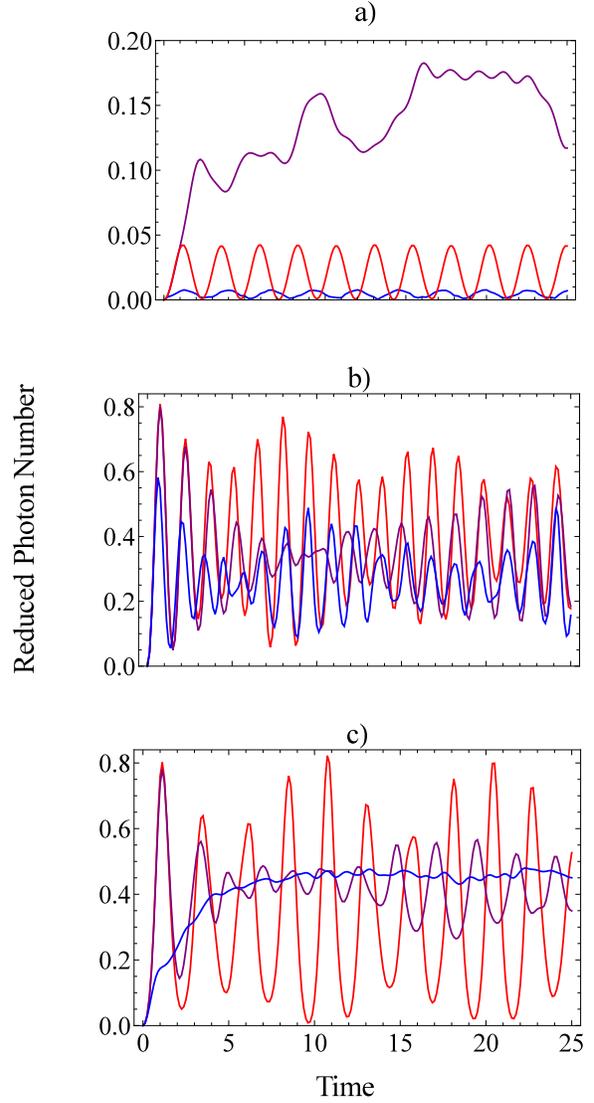}
	\caption{(Color online) 
	Long-term evolution of the reduced photon number $P_{\text{ph}}(t/\tau)$ ($\tau = 1/\Omega$) for 
	different couplings $J$ is shown. 
	In (a) and (b) the initial state is the ground state of
	the $XY$ chain for anti-ferromagnetic and ferromagnetic
	case, respectively. 
	The parameters used for the computations are $p=12,N=6$, $\Delta\tau = 0$, and $J\tau/\hbar = 0, 0.5, 3$ for (a) and $J\tau/\hbar = 0,-0.5,-3$ for (b). 
	In (c) the initial state is ferromagnetic with all spins($\mu$'s) up(down) $p=N=12$,
	the coupling changes in the range $J\tau/\hbar = 0,-0.5,-3$	and $\Delta\tau = 0$.
	The dimensionless time $t/\tau$ is plotted on the abscissa. 
	With the increase of absolute value of $J$ the colors change from red to blue.
	}
	\label{Fig:PhotonEmission}
\end{figure}

We consider some examples of the dynamics of the system 
that begins its evolution from
an initial state that contains all the excitations in the chain
($N\leq p$). 
We are interested in the on-resonance case ($\Delta = 0$) in order to obtain 
larger transfer of excitations from the chain to the field.
It is worth noticing that regardless of the arbitrary choice of
the initial state we always end up with the same behavior of the
physical quantities
-- irregular oscillations around some average value. 

The reduced photon number $P_{\text{ph}}(t)$ as a function of time and for 
different values of the coupling $J$ is shown in Fig. \ref{Fig:PhotonEmission}. 
Figures \ref{Fig:PhotonEmission} a) and \ref{Fig:PhotonEmission} b) represent the case when the initial state is 
the ground state of the $XY$ model for anti-ferromagnetic ($J>0$) and 
ferromagnetic $(J<0)$ cases, respectively, see Appendix \ref{XY_Diagonalization_22}. 
In Fig. \ref{Fig:PhotonEmission} c) the initial state is ferromagnetic with all spins up.
It is obvious that in the
regime $|J|\tau/\hbar \ll 1$ the dynamics corresponds mainly 
to that of the resonant Tavis-Cummings model.  
Most of the curves in Fig. \ref{Fig:PhotonEmission}
show an oscillatory behavior
with frequencies that are barely dependent on $J$.  
We observe a reduction of the amplitude of oscillations with the increase of $|J|$, 
since viewed in the $\eta$-fermion basis, 
$|J|$ increases the effective detuning. 
In general, further increase of $|J|$ would cause the excitations to 
stay predominantly trapped in the chain, 
and thence, decrease of the averaged over time value of
$P_{\mathrm{ph}}(t)$.
The latter effect is not clearly seen in the figures but is
additionally verified. 
These observations are very well illustrated in Fig. 7 c) where, for the given set of initial conditions, the number of excitations is $N = p =12$. 
In addition, the larger number of excitations used to obtain
the results in Fig.
\ref{Fig:PhotonEmission} c) in comparison with Figs. \ref{Fig:PhotonEmission} a) and \ref{Fig:PhotonEmission} b), 
leads to additional modes that are involved in the dynamics and more intensive exchange 
of excitations between the chain and the field. 
As a result the oscillations of $P_{\text{ph}}(t)$ for large $|J|$
practically vanish. 
It stabilizes around some average value where an equilibrium between 
the absorbed and radiated photons from the chain is established. 
See the curve corresponding to $J\tau/\hbar=-3$ in Fig.
\ref{Fig:PhotonEmission} c).

Another thing to notice is the short-time dynamics in Fig. \ref{Fig:PhotonEmission} c). 
More specifically, in the first period of oscillation the curves for 
$J\tau/\hbar=0,-0.5$ lie very close to each other. 
Therefore, since for the Tavis-Cummings model ($J=0$) the system
undergoes a superrradiant 
behavior during the emission ($m(t)\approx 0$,
$P_{\text{ph}}(t)\approx 1/2$). A similar behavior can be expected for
small values of
$J$ \footnote{For superradiance of a similar model system see also
\cite{Tokihiro1993}.}. In the second half-period the absorption in 
the "super" regime \cite{Higgins_superabsorption_2014} is realized as well. 
The increase of $J$ leads to the suppression of the oscillations involving super-processes.
In Figs. \ref{Fig:PhotonEmission} a) and \ref{Fig:PhotonEmission}  b) the magnetization $m(t)$ coincides with $P_{\text{ph}}(t)$, 
and $m(t)\approx 0$ is realized periodically in Fig. \ref{Fig:PhotonEmission} a) and only at the
beginning in Fig. \ref{Fig:PhotonEmission} b). 
In Fig. \ref{Fig:PhotonEmission} a) the emission rate, as seen, is expected to be small. 
In the first period of oscillation, in Fig. \ref{Fig:PhotonEmission} b), for $J\tau/\hbar=0,-0.5$,
however, we can assume that there is
a somewhat enhanced radiation. 
The above qualitative considerations are verified numerically by checking the condition 
\begin{align}
\left|\frac{I}{pI_0}\right| = \left|\frac{N (\text{d}P_{\text{ph}}(t)/\text{d}t)\vert_{\text{max}}}{pI_0}\right| > 1, 
\label{SuperradianceCond}
\end{align} 
which signals abnormal radiation/absorption.
Here $I_0$ and $I$ are the maximum radiation rates of a single spin and 
full system, respectively.
The condition (\ref{SuperradianceCond}) proves to be fulfilled
(compatible with the expected values for superradiant behavior) 
for the first period for the case of $J\tau/\hbar=0,-0.5$ 
in Fig. \ref{Fig:PhotonEmission} c), and slightly fulfilled for the first period 
for the case of $J\tau/\hbar=0,-0.5$ in Fig. \ref{Fig:PhotonEmission} b), as we have already discussed.

\section{Possible Applications}\label{Sec_PossibleApplications}

For the sake of clarity, here we
propose a couple of applications that can be described within the framework of
Hamiltonian \eqref{Hamiltonian_RWAt}. Furthermore we
estimate the corresponding model parameters.


\begin{figure*}[th!]
	\centering              
	\includegraphics[width=.9\textwidth]{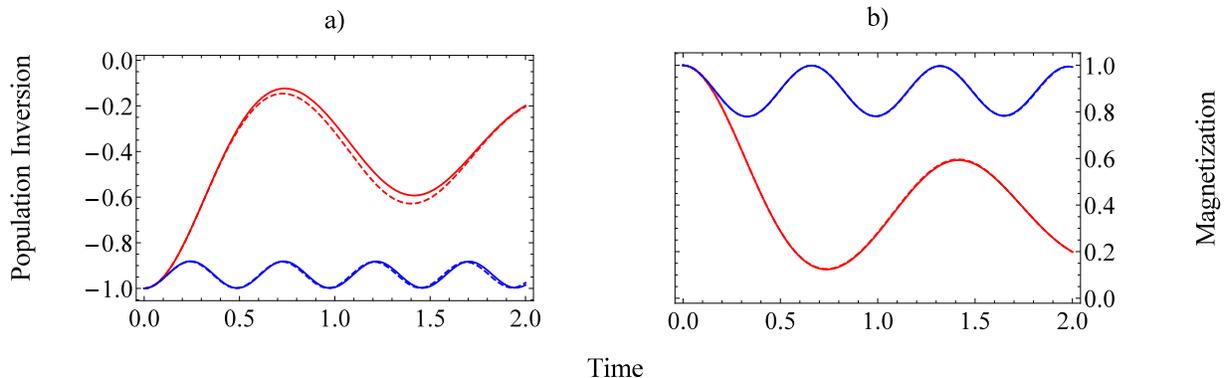}
	\caption{(Color online)  
		Dynamics of the averaged population inversion for $J$-aggregate (left panel) and dimensionless averaged magnetization for $XY$ spin-magnetic system (right panel) in cavity as functions of dimensionless time $t/\tau$ ($\tau = 1/\Omega$). 
		These are obtained with the use of the full Hamiltonian (\ref{Hamiltonian}) and the RWA-Hamiltonian (\ref{Hamiltonian_RWAt}), and are given with solid and dashed curves respectively.  
		Parameters used for the computations are $ p=10, \,
		N=4, \, \Omega =1/\tau \, = 10^{14}$ rad s$^{-1}$,
		$-J\tau/\hbar = 1, \, \omega_1\tau = 10, \, \omega\tau
		= 9.9 \,\,\text{and}\,\, 20 , \, \Delta\tau = 0.1
		\,\,\text{and}\,\, - 10 $, respectively for (a) and
		$ \Omega =1/\tau \, = 10^{9}$ rad s$^{-1}$,
		$-J\tau/\hbar = 1, \, \omega_1\tau = 10^2, \,
		\omega\tau = 99.9$
		and $90$ $\Delta\tau = 0.1$ and $10$, respectively for (b). 
		The curves corresponding to larger (smaller) by
		absolute values of the detununing are colored with blue (red). 
		In the full Hamiltonian diagonalization the state space of the left/right circularly polarized photons is truncated to six/three photons.  
	}
	\label{Fig:Applications}
\end{figure*}


\subsection{$J$-aggregate in a cavity }\label{SuBSec_MolecularAggregateInACavity}

The model of interest can potentially find application in describing certain type of 
linear molecular aggregates, namely the so called $J(H)$-aggregates \cite{Jelley1936,Sheibe1936,Kobayashi1996,Kobayashi2012}, 
coupled to a cavity mode. 
The Frenkel exciton model of these aggregates consists of one-dimensional ordered array of interacting identical two-level systems. 
It is assumed that the resonant electric dipole-dipole
interaction, say $J$, between excitons is restricted to
nearest neighbors only. 
The excitonic system is approximately modeled by the $XY$ chain, and here by assumption all other possible terms are corrections of higher order. 
The case when $J<0$ is known as $J$-aggregate and the case
of $J>0$ is referred to as $H$-aggregate.
Moreover, when the photon-exciton
interaction can be approximately described by the Tavis-Cummings
model, then the problem of $J(H)$-aggregate can be mapped onto
Hamiltonian (\ref{Hamiltonian_RWAt}) \footnote{Remark that this is an interaction between the electric dipole and the electric component of the field.}.

Consider a $J$-aggregate consisting of  $p \sim 10$ molecules, where the 
value of the binding energy of the Frenkel excitons is of
order $\hbar\omega_1 \sim 1\, eV$ 
($\omega_1 \sim \, 10^{15}$ rad s$^{-1}$), 
and the exciton-exciton and exciton-photon couplings are of order
$-J, \hbar\Omega \sim  0.1 \, eV$ ($|J|/\hbar,\Omega \sim \, 10^{14}$
rad s$^{-1}$). 
One can excite this aggregate by placing it
in a cavity supporting mode with frequency of order $\omega \sim
10^{15}$ rad s$^{-1}$.

A specific example for the averaged inversion $-m(t)$ for the near-resonance and far-off-resonance cases is given in Fig. \ref{Fig:Applications} a).
It is seen that the RWA is still valid, though conditions (\ref{RWA_CondSmallDModified}) are not fulfilled  strictly. 
The magnitudes of deviation from the ground state is significant for
near-resonance case, and large enough to be measured for
far-off-resonance case.
The time of observation of the oscillations is of
order $\tau \sim 10^{-14}\,$ s\footnote{A
rough estimate for the magnitude and period of oscillations can be
made by recalling the conjecture from Section
\ref{Sec_EffectiveHamiltonians} and using the solution
(\ref{Pm_SinglePhoton}) for a single-photon case \label{Apps_Footn}.}.

\subsection{Spin-magnetic $\boldsymbol{XY}$ chain in a cavity}\label{SubSec_Spin-MagneticChainInACavity}

Another possible application is to think literally of a spin-magnetic
$XY$ chain in a single-mode cavity. Consider such a chain consisting
of $p \sim 10$ spins placed in a magnetic field $B_z \sim 1$ T (hence
$\omega_1 \sim  10^{11}$ rad s$^{-1}$). We assume that its
(effective) spin-spin coupling and the spin-magnetic moment--photon
coupling are of the same order - $\,\, \hbar\Omega, - J \sim 10\, \mu
eV$ ($\Omega, |J|/\hbar\sim 10^{9}$ rad s$^{-1}$). In this case a
cavity whose mode has a frequency of order of $\omega \sim  10^{11}$
rad s$^{-1}$ would be relevant to excite the spin-magnetic system.

We use some exemplary values of experimental parameters, in order to
obtain the dynamics of the (dimensionless) averaged magnetization
$m(t)$ for both small and large detuning $\Delta$; These are shown in
Fig. \ref{Fig:Applications} b). In this case the conditions for the
validity of RWA are well fulfilled and both curves, presenting the
approximate and exact results, practically coincide.  The
demagnetization is large enough to be measured and the time for
observation of oscillations is of order $\tau \sim 10^{-9}$s $^{\ref{Apps_Footn}}$.

Note that if one wants to realize non-negligible excitations for
optical frequencies $\omega$, because of the very large detuning that
cannot be compensated with $\omega_1$ through the magnetic field
$B_z$, one needs to increase the spin-magnetic moment--photon coupling
by increasing the number of spins $p$ in the chain.
Then the validity of the RWA in this case may be questionable. 
Moreover, such an effect most probably will be shadowed by other indirect effects, except in the case when special measures are taken.

\section{Conclusions}\label{Sec_Conclusions}

We considered the model of a periodic $XY$ spin chain,
subject to a constant homogeneous magnetic field, 
and through a Dicke interaction to a single-mode of quantized
circularly polarized electromagnetic field.
For convenience, the external magnetic field and the $z$ axis of the
magnetic moments in the $XY$ chain are chosen to be collinear to the
cavity axis. 

The domain of validity of RWA applied to the Dicke
interaction component in the parameter space of the model is determined.
These are natural generalizations of
those corresponding to a single two-level system.
Assuming that all spins of the $XY$ chain interact with the field mode 
(thus excluding the field-boundary $XY$ term)
and the coupling between the spins is not much larger than 
the spin-magnetic moment--photon coupling (which is the case of interest)
the rest of the Hamiltonian is (approximately) kept unaffected by the RWA procedure.
Then, after application of successive Jordan-Wigner transformation and Fourier transform 
to the RWA Hamiltonian we continue the analysis in the
corresponding fermion basis, 
where the $XY$ component of the Hamiltonian is diagonal.
We introduced three observables to
describe the physics of the system: reduced
photon number, (dimensionless) magnetization per site and the ratio
between the total number of excitations and the number of spins. The
conservation of the total number of excitations relates them, so
only two quantities can be considered as independent.

We were mostly interested in the case when the initial state is a
product state of a number state of the field and the ground state of
the $XY$ model in the strong constant homogeneous field. We found the
exact analytical solution of the problem in the single-excitation
regime. We noticed that there are only two states involved in the
dynamics -- the state with the photon in the field, and the one with
the excited zero-momentum mode of the chain. Then the problem
effectively reduces to the solution of the well-known Jaynes-Cummings
model, which reveals the quantum Rabi oscillations.

For the general case of
multiple excitations, we performed numerical computations. 
It is shown that for the studied choices of parameters ($\Delta$
always large), the reduced photon number and the magnetization per
site also reveal a sort of oscillatory behavior with frequency
dependent on the values of $\Delta$ and $J$ for fixed $\Omega$, and
not so much on the values of $p$ and $N$. Slight dependence of
$P_{\text{ph}}(t)$ and strong dependence of $m(t)$ on $N$ is
established. Symmetry with respect to simultaneous changes
$P_{\text{ph}}(t)\leftrightarrow m(t)$ and $p\leftrightarrow 2N$ is
demonstrated. The increase of $|\Delta|$ is seen to lead to the
reduction of the amplitude and increase of the frequency of the
oscillations.

Based on the assumption that the vast part of the excitations are shared between the field 
and the zero-momentum modes of the chain we found two effective Hamiltonians. 
The first is an effective matrix tridiagonal Hamiltonian acting on the space spanned by the 
modes possessing zero net momentum, which have the smallest (in absolute value) set of 
momenta for a given number of excitations in the field.
The second is the Tavis-Cummings effective Hamiltonian, 
found by smearing out the difference between even and odd subspaces and 
setting all the momenta to zero. 
The former (tridiagonal) Hamiltonian acts on the much reduced in size Hilbert subspace, 
that leads to a great simplification of the computations.
The latter (Tavis-Cummings) Hamiltonian has the advantage that it is a well-known 
and largely studied model.
It is demonstrated that the effective Hamiltonians generate short-time dynamics 
that provides a reasonable approximation for small number of total excitations.

The on-resonance situations, when the initial condition is such that 
the excitations reside in the $XY$ chain, are also discussed. 
It is found that the long-time evolution brings the 
system, generally, in an oscillatory regime around some 
average value of the reduced photon number. 
The tendencies of the amplitude of the oscillations and 
of the average value of $P_{\text{ph}}(t)$ is to decrease with the increase of 
the interaction $|J|$ between spins, 
and for larger $|J|$ the amplitude of the oscillations decreases with the 
increase of the number of excitations $N$.
A specific example is considered, 
with maximally excited $XY$ chain component of the initial condition, and for small $|J|$,
where the superradiant behavior shows up at the beginning of the emission.  

Generally speaking, the results presented here can be useful in studying problems such as 
bosonic mode in a spin-bath, spin chains or linear molecular aggregates 
in a single-mode cavity, and superradiating/superabsorbing systems.
Specific applications, namely spin chain and $J$-aggregate interacting
with a single cavity mode, with specific
values for the experimental parameters, are shortly discussed.

\acknowledgments

The authors would like to thank A. A. Donkov for useful discussions
during an early stage of this project.
S. Varbev is supported by the Bulgarian Ministry of Education and Science under the National
Research Programme ``Young scientists and postdoctoral researchers''
approved by DCM under No 577/17.08.2018.
I. Boradjiev, H. Tonchev and H. Chamati would like to acknowledge the financial support by Grants
DN08/18 and KP-06-N38/6 of the Bulgarian National Science Fund.

\appendix
\section{Rotating wave approximation} \label{RWA}

In this Appendix we present the derivation of Hamiltonian
\eqref{Hamiltonian_RWA}
following the approach presented in \cite{Puri2001}.
To this end, we make use of the formula
\begin{subequations} \label{TExpFormula}
\begin{equation}
\overleftarrow{T}\exp \left\{\int_{t_0}^t \d\tau \left[\hat{A} + \hat{B}(\tau)\right]\right\} =
\exp\left[\hat{A} (t-t_0)\right] \overleftarrow{T} \exp \left[ \int_{t_0}^t \d\tau \,\, \hat{\bar{B}}(\tau) \right]  , 
\end{equation}
where
\begin{equation}
\hat{\bar{B}}(t) = \exp\left[- \hat{A} (t-t_0)\right] \hat{B} \,\, \exp\left[ \hat{A} (t-t_0)\right],
\end{equation}
and $\overleftarrow{T}$ stands for the time-ordering of the exponent.
\end{subequations}

The formal solution of the Schr\"{o}dinger equation 
\begin{align}
\i\hbar \, \frac{\d}{\d t} \vert\psi\rangle = \hat{H} \vert \psi\rangle, \quad
\vert\psi_0\rangle = \vert\psi(0)\rangle , 
\end{align}
reads
\begin{equation}
\vert \psi\rangle = \overleftarrow{T}\exp\left(-\frac{\i}{\hbar}
	\int_{0}^t \d\tau \,\, \hat{H} \right)  \vert \psi_0\rangle.
\end{equation}

Identifying in \eqref{Hamiltonian} the operators $\hat{A}$ with
$-\i\omega\left(\hat{n}_+ + \hat{n}_-  +
\frac{p}{2} + \sum_{i=1}^p \hat{S}_{iz}\right)$ and $\hat{B}$ with
$-\tfrac i\hbar (\hat{H}^{XY} + \hat{H}^D)$, respectively, we get
\begin{align}
	\vert \psi\rangle = & \exp\left(-\i\omega\left(\hat{n}_+ + \hat{n}_-  +
	\frac{p}{2} + \sum_{i=1}^p \hat{S}_{iz}\right) T \right) \notag \\
	& \qquad \times \overleftarrow{T}\exp\left(-\frac{\i}{\hbar} \int_{0}^T \text{d}t \,\, \hat{\bar{H}}^{XY+D} \right) \vert \psi_0\rangle, \label{psi_N}
\end{align}
where 
\begin{subequations}
\begin{align}
	\hat{\bar{H}}^{XY+D} = \ & \hat{H}^{XY}_{>p}
	+\hat{H}^{XY}_{<p} 
	+  \hbar \omega_1 \,\sum_{i>p} \hat{S}_{iz} +\hat{H}^{RWA}
	\notag \\
	& \ + J \sum_{i > p \geq j} \left(\hat{S}_{i+} \hat{S}_{j-}
	\e^{-\i\omega t} + \hat{S}_{i-} \hat{S}_{j+} \e^{+\i\omega t}
	\right) \notag \\
	& \ - \frac{\hbar\Omega}{\sqrt{2}} \sum_{i=1}^p \left( \hat{a}_-^{\dagger} \hat{S}_{i+} \e^{2\i\omega t} + \hat{a}_- \hat{S}_{i-} \e^{-2\i\omega t} \right),
\end{align}
with
\begin{equation}
\hat{H}^{XY}_{>p} =  J \sum_{i > j > p} \left(\hat{S}_{i+} \hat{S}_{j-} + \hat{S}_{i-} \hat{S}_{j+} \right) ,
\end{equation}
\begin{equation}
\hat{H}^{XY}_{<p} = J \sum_{p \geq i > j} \left(\hat{S}_{i+} \hat{S}_{j-} + \hat{S}_{i-} \hat{S}_{j+} \right) ,
\end{equation}
and
\begin{equation}
\hat{H}^{RWA} = \hbar \Delta \,\sum_{i=1}^p \hat{S}_{iz} 
    - \frac{\hbar\Omega}{\sqrt{2}} \sum_{i=1}^p \left( \hat{a}_+ \hat{S}_{i+}  +  \hat{a}_+^{\dagger} \hat{S}_{i-} \right)  . \label{H_RWA_S}
\end{equation}
\end{subequations}

Now we apply formula (\ref{TExpFormula}) for the time-ordered exponent
in (\ref{psi_N}), with the identification $\hat{A} = -(\i/\hbar)
\hat{H}^{RWA}$ and $\hat{B} = -(\i/\hbar) (\hat{\bar{H}}^{XY+D} -
\hat{H}^{RWA})$.
\begin{subequations}
\begin{align}
	& \overleftarrow{T}\exp\left(-\frac{\i}{\hbar} \int_{0}^T
	\text{d}t \,\, \hat{\bar{H}}^{XY+D} \right) = \exp\left(-\frac{\i}{\hbar} \, \hat{H}^{RWA} \, T\right) \notag \\
	& \times \overleftarrow{T} \exp \left(-\frac{\i}{\hbar} \int_{0}^T \text{d}t \,\, \hat{\bar{H}}^{XY+D-RWA} \right), \label{TExp_XY-N}
\end{align}
with
\begin{align}
	& \hat{\bar{H}}^{XY+D-RWA} = \hat{H}^{XY}_{>p} + \hat{\bar{H}}^{XY}_{<p} + 
	\hbar \omega_1 \,\sum_{i>p} \hat{S}_{iz} \notag \\
	& \qquad + J \sum_{i > p \geq j} \left[ \hat{S}_{i+} \hat{S}_{j-}(t)
	\e^{-\i\omega t} + \hat{S}_{i-} \hat{S}_{j+}(t) \e^{+\i\omega
	t} \right] \notag \\
	& \qquad - \frac{\hbar\alpha_-\Omega}{\sqrt{2}} \sum_{i=1}^p \left[
	\hat{S}_{i+}(t) \e^{2\i\omega t} + \hat{S}_{i-} (t)
	\e^{-2\i\omega t} \right]  ,
\end{align}
and
\begin{equation}
\hat{\bar{H}}^{XY}_{<p} = J \sum_{p \geq i > j} \left[\hat{S}_{i+} (t) \hat{S}_{j-} (t) + \hat{S}_{i-} (t) \hat{S}_{j+} (t) \right] . \label{H_XY_bar_Lp}
\end{equation}
\end{subequations}
To derive the above equation, we proceed as follows:
We introduce a new set of half-spin operators
\begin{align}
 \hat{S}_{i+} &= \frac{2\Gamma}{\Lambda} \,  \hat{R}_{iz} 
 + \frac{\i}{2} \, \left(1+\frac{\delta}{\Lambda}\right) \,  \hat{R}_{i+} + 
 \frac{\i}{2} \, \left(1-\frac{\delta}{\Lambda}\right) \,  \hat{R}_{i-}, \notag \\
 \hat{S}_{i-} &= \frac{2\Gamma}{\Lambda} \,  \hat{R}_{iz} 
 - \frac{\i}{2} \, \left(1-\frac{\delta}{\Lambda}\right) \,  \hat{R}_{i+} - 
 \frac{\i}{2} \, \left(1+\frac{\delta}{\Lambda}\right) \,  \hat{R}_{i-}, \notag \\
 \hat{S}_{iz} &= \frac{\delta}{\Lambda} \,  \hat{R}_{iz} 
 - \frac{\i\Gamma}{\Lambda} \, \left(\hat{R}_{i+} - \hat{R}_{i-}\right) , 
 \quad i=1,\dots, p,
\end{align}
where 
\begin{align}
[\hat{R}_{i+},\hat{R}_{i-}] = 2\hat{R}_{iz}, \quad 
[\hat{R}_{iz},\hat{R}_{i\pm}] = \pm\hat{R}_{i\pm}.
\end{align}
We consider the case of a large photon number, so we make the
substitution
\begin{align}
\hat{a}_{\pm}, \, \hat{a}^{\dagger}_{\pm} \rightarrow \alpha_{\pm}.
\end{align}
Further, we define the remaining free constants that parameterize the transformation
\begin{align}
\delta = \Delta, \qquad \sqrt{2}\Gamma = - \alpha_{+}\Omega, \qquad \Lambda=\sqrt{4\Gamma^2+\delta^2}.
\end{align}
In this representation we have
\begin{align}
\hat{H}^{RWA} = \hbar\Lambda \sum_{i=1}^{p}\hat{R}_{iz}. 
\end{align}
The operators $\hat{S}_{i\pm}(t)$ defined via  
\begin{equation}
 \hat{S}_{i\pm}(t) = \exp\left(\frac{\i}{\hbar} \, \hat{H}^{RWA} \, t\right)
 \hat{S}_{i\pm} \exp\left(- \frac{\i}{\hbar} \, \hat{H}^{RWA} \, t\right) , \label{S_t}
\end{equation}
are easy to calculate explicitly with the use of the formulas
\begin{equation}
\exp\left(\i\Lambda t \hat{R}_{iz}\right) \, \hat{R}_{i\pm} \exp\left(-\i\Lambda t \hat{R}_{iz}\right) = 
\exp\left(\pm\i\Lambda t\right) \, \hat{R}_{i\pm}.
\end{equation}
Next, we apply formula (\ref{TExpFormula}) for the time-ordered
exponent in the right-hand side of (\ref{TExp_XY-N}), with the
identification $\hat{A} = -(\i \omega_1 \,\sum_{i>p} \hat{S}_{iz}) $
and $\hat{B} = -(\i/\hbar) (\hat{\bar{H}}^{XY+D-RWA} - \hbar \omega_1
\,\sum_{i>p} \hat{S}_{iz} )$, 
\begin{align}
	& \overleftarrow{T}\exp\left(-\frac{\i}{\hbar} \int_{0}^T
	\text{d}t \,\, \hat{\bar{H}}^{XY+D-RWA} \right) \notag \\
	& \quad = \exp\left(-\i \omega_1 \,\sum_{i>p} \hat{S}_{iz}  \, T\right) \overleftarrow{T} \exp \left( -\frac{\i}{\hbar} \int_{0}^T \text{d}t \,\, \hat{\bar{H}}^{XY+D-RWA-Z } \right),  \label{TExpXY+D-RWA}
\end{align}
with
\begin{align}
	& \hat{\bar{H}}^{XY+D-RWA-Z}
	=  \hat{H}^{XY}_{>p} + \hat{\bar{H}}^{XY}_{<p} \notag \\
	& \qquad  + J \sum_{i > p \geq j} \left[ \hat{S}_{i+}
	\hat{S}_{j-}(t) \e^{+\i\Delta t} + \hat{S}_{i-}
	\hat{S}_{j+}(t) \e^{-\i\Delta t} \right] \notag \\
	& \qquad - \frac{\hbar\alpha_-\Omega}{\sqrt{2}} \sum_{i=1}^p \left[ \hat{S}_{i+}(t) \e^{2\i\omega t} + \hat{S}_{i-} (t) \e^{-2\i\omega t} \right].
\end{align}

The application of formula (\ref{TExpFormula}) for the
time-ordered exponent in the right-hand side of (\ref{TExpXY+D-RWA}),
with the identification $\hat{A} = -(\i/\hbar) \hat{H}^{XY}_{>p}$ and
$\hat{B} = -(\i/\hbar) (\hat{\bar{H}}^{XY+D-RWA-Z} - \hat{H}^{XY}_{>p}
)$, gives
\begin{subequations}
\begin{widetext}
\begin{equation}
\overleftarrow{T}\exp\left(-\frac{\i}{\hbar} \int_{0}^T \text{d}t
	\,\, \hat{\bar{H}}^{XY+D-RWA-Z} \right) =
	\exp\left(-\frac{\i}{\hbar} \, \hat{H}^{XY}_{>p} \, T\right)
	\overleftarrow{T} \exp \left( -\frac{\i}{\hbar} \int_{0}^T
	\text{d}t \,\, \hat{\bar{H}}^{XY+D-RWA-Z-(XY>p)} \right),
	\label{TExpXY+D-RWA-Z}
\end{equation}
with
\begin{align}
\hat{\bar{H}}^{XY+D-RWA-Z-(XY>p)}
	= \ & \hat{\bar{H}}^{XY}_{<p} +  J \sum_{i > p \geq j} \left[
		\hat{S}_{i+}^{\prime}(t) \hat{S}_{j-}(t) \e^{+\i\Delta
		t} + \hat{S}_{i-}^{\prime}(t) \hat{S}_{j+}(t)
		\e^{-\i\Delta t} \right] \notag \\
	& - \frac{\hbar\alpha_-\Omega}{\sqrt{2}} \sum_{i=1}^p \left[ \hat{S}_{i+}(t) \e^{2\i\omega t} + \hat{S}_{i-} (t) \e^{-2\i\omega t} \right],
\end{align}
\end{widetext}
where
\begin{align}
	\hat{S}_{i\pm}^{\prime}(t) & = \exp\left(+\frac{\i}{\hbar}
	\, \hat{H}^{XY}_{>p} \, t\right) \hat{S}_{i\pm}
	\exp\left(-\frac{\i}{\hbar} \, \hat{H}^{XY}_{>p} \, t\right) \label{S_Prime_t} .
\end{align}
\end{subequations}
We notice that $\hat{S}_{i\pm}^{\prime}(t)$ are still half-spin operators. 
This can be explicitly seen by first changing to the $\eta$-representation, simplifying the expression and redefining the phases of $\hat{\eta}$ operators to the relevant time-dependent ones, 
then moving back to modified $c(t)$-representation, and eventually to
$S^{\prime}(t)$-representation. The respective notations and formulas
are given in Appendix \ref{XY_Diagonalization}.

From now on we work with one-dimensional chain in nearest-neighbor
approximation. In this case the sum involving
$\hat{S}_{i\pm}^{\prime}(t)$ is reduced to a single term ($i=p+1$).
Again, the application of the formula (\ref{TExpFormula}) for the
time-ordered exponent in the right-hand side of
(\ref{TExpXY+D-RWA-Z}), with the identification $\hat{A} = -(\i/\hbar)
\hat{H}^{XY}_{<p}$ and $\hat{B} = -(\i/\hbar)
(\hat{\bar{H}}^{XY+D-RWA-Z-(XY>p)} - \hat{H}^{XY}_{<p} )$, gives
\begin{subequations}
\begin{widetext}
\begin{equation}
\overleftarrow{T}\exp\left(-\frac{\i}{\hbar} \int_{0}^T \text{d}t \,\,
	\hat{\bar{H}}^{XY+D-RWA-Z-(XY>p)} \right) =
	\exp\left(-\frac{\i}{\hbar} \, \hat{H}^{XY}_{<p} \, T\right)
	\overleftarrow{T} \exp \left( -\frac{\i}{\hbar} \int_{0}^T
	\text{d}t \,\, \hat{\bar{H}}^{XY+D-RWA-Z-(XY>p)-(XY<p)}
	\right),
\end{equation}
with
\begin{align}
\hat{\bar{H}}^{XY+D-RWA-Z-(XY>p)-(XY<p)} = &
J \sum_{i > p \geq j} \left[ \hat{S}_{i+}^{\prime}(t)
	\hat{S}^{\prime\prime}_{j-}(t) \e^{+\i\Delta t} +
	\hat{S}_{i-}^{\prime}(t) \hat{S}^{\prime\prime}_{j+}(t)
	\e^{-\i\Delta t} \right] \notag \\
	& -
	\frac{\hbar\alpha_-\Omega}{\sqrt{2}} \sum_{i=1}^p \left[
	\hat{S}^{\prime\prime}_{i+}(t) \e^{2\i\omega t} +
	\hat{S}^{\prime\prime}_{i-} (t) \e^{-2\i\omega t}
	\right] + \hat{\bar{H}}^{\prime \, XY}_{<p} -\hat{H}^{XY}_{<p},
\end{align}
\end{widetext}
and
\begin{align}
	\hat{S}_{i\pm}^{\prime\prime}(t) & =
	\exp\left(+\frac{\i}{\hbar} \, \hat{H}^{XY}_{<p} \, t\right)
	\hat{S}_{i\pm}(t) \exp\left(-\frac{\i}{\hbar} \,
	\hat{H}^{XY}_{<p} \, t\right) \notag \\
	& =\exp\left(+\frac{\i}{\hbar} \, \hat{H}^{\prime\,RWA}(t) \, t\right)
	\hat{S}_{i\pm}^{\prime}(t) \exp\left(-\frac{\i}{\hbar} \,
	\hat{H}^{\prime\,RWA}(t) \, t\right) ,
\end{align}
\end{subequations}
where $\hat{\bar{H}}^{\prime \, XY}_{<p}$ is given via (\ref{H_XY_bar_Lp}) but with $\hat{S}_{i\pm}(t)$ replaced by $\hat{S}_{i\pm}^{\prime\prime}(t)$, 
$\hat{H}^{\prime\,RWA}(t)$ is given via (\ref{H_RWA_S}) but with
$\hat{S}_{i\pm}$ replaced by $\hat{S}_{i\pm}^{\prime}(t)$, and
$\hat{S}_{i\pm}^{\prime}(t)$ is the analog of (\ref{S_Prime_t}) but
for the component of the chain in the cavity i.e. for $i\leq p$. 
$\hat{S}_{i\pm}^{\prime\prime}(t)$ can be calculated explicitly, in terms of some operators $\hat{R}(t)$, as it is done for $\hat{S}(t)$ (\ref{S_t}).
We point in advance that, when we derive the RWA conditions, we will use the following estimates 
$\hat{R}_{iz}(t) \rightarrow 1/2$, $\hat{R}_{i\pm}(t) \rightarrow 1$, 
and $\hat{S}_{i\pm}^{\prime}(t) \rightarrow 1$.

Strictly speaking, for the validity of RWA, the frequency $\omega$ must be 
orders of magnitude larger than any other quantity. 
To employ the modified for the case RWA is equivalent to approximate the $T$-exponent 
\begin{widetext}
\begin{align}
\overleftarrow{T} \exp \left( -\frac{\i}{\hbar} \int_{0}^T \text{d}t \,\, \hat{\bar{H}}^{XY+D-RWA-Z-(XY>p)-(XY<p)} \right)
\approx \ & \overleftarrow{T} \exp \left\{
-\frac{\i}{\hbar} \int_{0}^T \text{d}t \,\, \left[ 
\left(-\frac{\hbar\alpha_-\Omega}{\sqrt{2}}\right) \sum_{i=1}^p \left(
	\hat{S}_{i+}^{\prime\prime}(t) \e^{2\i\omega t} + \hat{S}_{i-}^{\prime\prime} (t)
	\e^{-2\i\omega t} \right) \right. \right. \notag \\
	& \ + \left. \left.
J \sum_{i > p \geq j} \left( \hat{S}_{i+}^{\prime}(t) \hat{S}_{j-}^{\prime\prime}(t) \e^{+\i\Delta t} + \hat{S}_{i-}^{\prime}(t) \hat{S}_{j+}^{\prime\prime}(t) \e^{-\i\Delta t} \right) \right. \right. \notag \\
& \ \left.\left. + J \sum_{ p >i \geq j} \left( \hat{S}_{i+}^{\prime\prime}(t) \hat{S}_{j-}^{\prime\prime}(t) + \hat{S}_{i-}^{\prime\prime}(t) \hat{S}_{j+}^{\prime\prime}(t) \right) - J \sum_{ p >i \geq j} \left( \hat{S}_{i+} \hat{S}_{j-} + \hat{S}_{i-} \hat{S}_{j+} \right) \right] \right\} \label{TExpRWA}
\end{align}
\end{widetext}
with the unity operator.
To this end, it is sufficient to analyze the following 
expressions entering the exponent in the r.h.s. of (\ref{TExpRWA}) as summands. 
\begin{enumerate}[label=(\roman*)]
\item 
Expressions for the Dicke component (first term in the rhs of (\ref{TExpRWA})):
\begin{align}
1)&\quad  I_1=\int_0^{T} \text{d}t \, \cos(2\omega t)=\frac{\sin(2\omega T)}{2\omega} , \\
&\left|\i\sqrt{2}p\frac{\alpha_-\Omega\Gamma}{\Lambda}I_1 \right| \lesssim 
\begin{cases}
p\left|\frac{\alpha_-\Omega}{2\omega}\right|, 
& \text{for} \:\:\Delta \to 0,\\
p \left|\frac{\alpha_-\Omega}{2\omega}\right| \left|\frac{\alpha_+\Omega}{\omega}\right|  , 
& \text{for} \:\:\Delta \to -\omega.
\end{cases}
\end{align}
\begin{align}
2)&\quad (I^{-}_2)^{*}= I^{+}_2=\int_0^{T} \text{d}t \, e^{ \i\Lambda t}\left( \i\frac{\delta}{\Lambda}\cos(2\omega t)-\sin(2\omega t)\right)= \notag \\
&= \frac{2\omega - \delta + e^{\i\Lambda T}\left[ \i\sin(2\omega T)\left( \Lambda-2\omega\frac{\delta}{\Lambda}\right) - \cos(2\omega T)(2\omega-\delta) \right]}{\Lambda^2-4\omega^{2}},\\
&\quad  \left| \i p \frac{\alpha_-\Omega}{\sqrt{2}} I^{\pm}_{2} \right| \lesssim 
\begin{cases}
p\left|\frac{\alpha_-\Omega}{\omega}\right| , 
& \text{for} \:\:\Delta \to 0,\\
3p\left|\frac{\alpha_-\Omega}{\omega}\right| , 
& \:\: \text{for} \:\:\Delta \to -\omega.
\end{cases}
\end{align}
\item 
Expressions for the field-boundary $XY$ term
\footnote{Hereafter we will refer to this term as to a field-boundary $XY$ term 
in the sense that it describes the coupling between the interacting and noninteracting 
with the field mode parts of the $XY$ chain.  
It must not be confused with the boundary terms defined by the ends of the chain.},
(second term in the rhs of (\ref{TExpRWA})):
\begin{align}
3) \quad &(I^{-}_3)^{*}=I^{+}_3=\int_0^{T} \text{d}t \, e^{ \i\Delta t}= -\i\frac{e^{ \i\Delta T}-1}{\Delta} , \\
&\left|-\i \frac{J\Gamma}{\hbar\Lambda}I^{\pm}_3\right| 
\lesssim
\begin{cases}
\left| \frac{JT}{2\hbar}\right| , 
& \text{for} \:\:\Delta \to 0 \:\:\footnotemark,\\
\left| \frac{2J}{\hbar \omega} \right| \left| \frac{\alpha_+\Omega}{ \omega} \right|, 
& \text{for} \:\:\Delta \to -\omega.
\end{cases}
\end{align}
\setcounter{footnote}{9}
\footnotetext{This is nonnegligible in the case near resonance.}
\begin{align}
4) &\quad (I^{-}_{4a})^{*}=I^{+}_{4a}=-\i\int_0^{T} \text{d}t \, e^{\i(\Lambda + \Delta)t}= -\frac{e^{\i(\Lambda + \Delta) T}-1}{\Lambda + \Delta}, \\
&\qquad  (I^{-}_{4b})^{*}=I^{+}_{4b}= \i\int_0^{T} \text{d}t \, e^{\i(\Lambda - \Delta)t}= \frac{e^{\i(\Lambda - \Delta) T}-1}{\Lambda - \Delta}, \\
&\left| - i \frac{J}{2\hbar}\left(1 - \frac{\delta}{\Lambda}\right)I^{\pm}_{4a}\right| \lesssim 
\begin{cases}
 \left| \frac{J}{\hbar\alpha_+\Omega} \right| , 
 & \text{for} \:\:\Delta \to 0,\\
\left| 2\frac{J}{\hbar \omega} \left(\frac{\omega}{\alpha_+\Omega}\right)^2 \right|, 
& \text{for} \:\:\Delta \to -\omega \:\: \footnotemark,
\end{cases}\\
&\left| - i\frac{J}{2\hbar}\left(1 + \frac{\delta}{\Lambda}\right)I^{\pm}_{4b}\right| \lesssim 
\begin{cases}
 \left| \frac{J}{\hbar\alpha_+\Omega} \right|  , 
& \text{for} \:\:\Delta \to 0,\\
\left| \frac{1}{2}\frac{J}{\hbar \omega} \left(\frac{\alpha_+\Omega}{\omega}\right)^2 \right|, 
& \text{for} \:\:\Delta \to -\omega.
\end{cases}
\end{align}
\setcounter{footnote}{10}
\footnotetext{This is nonnegligible in the case of large $\omega$.} 
\item 
Expressions for the $XY$ component interacting with the field mode (last two terms in the rhs of (\ref{TExpRWA})):
\begin{align}
5) &\quad (I^{-}_5)^{*}=I^{+}_5=\i\int_0^{T} \text{d}t \, (e^{\i\Lambda t}-1) = \frac{e^{\i \Lambda T}-1}{\Lambda} -\i T, \\
& \left|-\i p \frac{J\Gamma\delta}{\hbar\Lambda^{2}} I^{\pm}_5 \right| \lesssim
\begin{cases}
p\left|\frac{J\Delta T}{2\hbar\alpha_+\Omega}\right|\left| \frac{2}{\alpha_{+}\Omega T}+1 \right|\to 0 , 
& \text{for} \:\: \Delta \to 0,\\
2p\left|\frac{J}{\hbar\omega}\right| \left|\frac{\alpha_+\Omega}{\omega}\right|, 
& \text{for} \:\: \Delta \to -\omega.
\end{cases}\\
6) &\quad (I^{-}_6)^{*}=I^{+}_6=\int_0^{T} \text{d}t \, (e^{2\i\Lambda t}-1) = -\i\frac{e^{2\i \Lambda T}-1}{2\Lambda} - T, \\
& \left|-i p\frac{J}{2\hbar}\left( 1-\frac{\delta^2}{\Lambda^2} \right) I^{\pm}_6\right| \lesssim
\begin{cases}
p\left|\frac{JT}{2\hbar}\right| \left| \frac{1}{\alpha_{+}\Omega T}+1 \right|,  
& \text{for} \:\:\Delta \to 0 \:\: \footnotemark,\\
p\left|\frac{J}{\hbar\omega} \left(\frac{\alpha_+\Omega}{\omega}\right)^{2} \right|, 
& \text{for} \:\:\Delta \to -\omega.
\end{cases}
\end{align}
\setcounter{footnote}{11}
\footnotetext{This is true when $(\alpha_{+}\Omega T)^{-1}$ is not very large.}
\end{enumerate}

From all of the above estimates we can conclude that:
\begin{enumerate} [label=(\roman*)]
\item 
Executing RWA for the Dicke Hamiltonian
is possible given that 
\begin{align}
\frac{\alpha_{+}\Omega}{\omega} \ll 1, \quad
p\,\frac{\alpha_-\Omega}{\omega} \ll 1 . \label{RWA_CondD} 
\end{align} 
\item
In general, the RWA will alter the field-boundary $XY$ term and it will be non-negligible. 
We have to recall, however, that the boundary term is of order $O(p^{-1})$, 
and so, for a very specific choice of parameters it may be still considered small.
To avoid possible complications from this term when performing RWA we will eliminate it. 
This can be achieved simply by considering an $XY$ chain 
coupled to a photon mode with all of its spins.
\item
Sufficient conditions to keep the interacting with the field mode $XY$ terms unchanged under the RWA read as follows. 
In the near-resonant case ($\Delta\to 0$), we require that 
\begin{align}
p \, \frac{JT}{\hbar} \ll 1 , \label{RWA_CondSmallD} 
\end{align}
and in the large detuning case ($\Delta\to -\omega$), we require
\begin{align}
p \, \frac{J}{\hbar\omega} \lesssim 1 . \label{RWA_CondBigD} 
\end{align}
However, the condition (\ref{RWA_CondSmallD}) can be relaxed for $J/\hbar \lesssim \Omega \lesssim \omega$. 
For large $J, J/\hbar\, \ll \Omega$, the correct approach would be to
		consider the whole Dicke component of the Hamiltonian as a perturbation.
\end{enumerate}

\section{Diagonalization of $\boldsymbol{XY}$ component of the
Hamiltonian. Ground state of the $\boldsymbol{XY}$ model in a strong constant homogeneous field} \label{XY_Diagonalization}

In this Appendix we briefly describe the transformations that
diagonalize the $XY$ component of Hamiltonian
(\ref{Hamiltonian_RWAt}) and find
its ground states and corresponding energies.

\subsection{Diagonalization of $\boldsymbol{XY}$ component of the Hamiltonian} \label{XY_Diagonalization_1}

We consider the periodic $XY$ model consisting of even number of spins in the nearest-neighbor approximation. 

Note that in the case of periodic boundary conditions, 
\begin{align}
\hat{S}_{p+1\pm} = \hat{S}_{1\pm}, \label{Cyclic_Boundary_Condition}
\end{align}
we have to add to the $XY$ component of the Hamiltonian (\ref{Hamiltonian_RWA}) the term 
\begin{align}
J \left(\hat{S}_{1+} \hat{S}_{p-} + \hat{S}_{1-} \hat{S}_{p+}\right) .
\end{align}

To begin with, we employ the Jordan-Wigner transformation, i.e. we introduce a set of spinless Fermi operators
\begin{align*}
\hat{S}_{i-} &= \exp\left( -\i\pi \sum_{j = 1}^{i-1}
	\hat{c}_{j}^{\dagger}\hat{c}_{j} \right) \hat{c}_{i}, \\
\hat{S}_{i+} & = \hat{c}_{i}^{\dagger} \exp\left( \i\pi \sum_{j = 1}^{i-1} \hat{c}_{j}^{\dagger}\hat{c}_{j} \right),
\end{align*}
obeying the canonical anti-commutation relations 
\begin{align}
\{\hat{c}_i, \hat{c}_j^{\dagger}\} = \delta_{ij}, \quad
\{\hat{c}_i, \hat{c}_j\} = 
\{\hat{c}_i^{\dagger}, \hat{c}_j^{\dagger}\}  = 0.
\end{align}
In terms of $\hat{c}$ and $\hat{c}^{\dagger}$ the expression for the Hamiltonian (\ref{Hamiltonian_RWA}) becomes \footnote{Here we skip the number operator for the sake of brevity.}
\begin{subequations}
\begin{equation}
\hat{H} = \hat{H}_p + \hat{H}_i + \hat{H}_b 
, \label{H_c+i+b} \\
\end{equation}
with
\begin{align}
\hat{H}_p =& \,\, J \sum_{i=1}^{p-1} \left( \hat{c}_{i+1}^{\dagger} \hat{c}_{i} + \hat{c}_{i}^{\dagger} \hat{c}_{i+1}  \right) + J \left( \hat{c}_{1}^{\dagger} \hat{c}_{p} + \hat{c}_{p}^{\dagger} \hat{c}_{1}  \right)\notag \\ 
&+  \hbar \Delta \,\sum_{i=1}^p \left(\hat{c}_i^{\dagger}\hat{c}_i - \frac{1}{2}\right) , \label{H_c}  \\
\hat{H}_i = &  \, - \frac{\hbar\Omega}{\sqrt{2}} \sum_{i=1}^p \left[ \hat{a}_+
\hat{c}_{i}^{\dagger} \exp \left(\i \, \pi \sum_{j = 1}^{i-1} 
\hat{c}_{j}^{\dagger}\hat{c}_{j}  \right)  + h.c. 
\right] ,  \\ 
\hat{H}_b =& -J \left(\hat{c}_1^{\dagger}\hat{c}_p +  \hat{c}_p^{\dagger}\hat{c}_1\right)\left(P + 1\right), \label{H_b}
\end{align}
\end{subequations}
where $P$ is the fermion parity operator
\begin{align}
P =\exp\left( \i\pi {\cal N}\right), \quad
{\cal N} = \sum_{j=1}^{p} \hat{c}_j^{\dagger}\hat{c}_j .
\end{align}
Observe that $P$ is a constant of motion with respect to $\hat{H}_p$ as
\begin{align}
[P,\hat{H}_p] = 0 ,
\end{align}
and accepts two values -- $P=+1$ for even and $P=-1$ for odd number of
fermions, respectively.
Therefore, the periodic $XY$ component $\hat{H}_p$ can be separately diagonalized in two subspaces with different parities, labeled by $\sigma$, see for example  \cite{DePasquale2008,Tokihiro1993,Wu2018}.
The identification is as follows: $\sigma = +1$ corresponds to a $c$-fermion system with imposed anti-periodic boundary conditions ($\hat{H}_b \neq 0$) and even number of excitations, while $\sigma = -1$ corresponds to the one with periodic boundary conditions ($\hat{H}_b = 0$) and with odd number of excitations.

Further, we rewrite the off-diagonal component of $\hat{H}_p+\hat{H}_b$, as
\begin{subequations}
\begin{equation}
2 J \sum_{i,j=1}^{p} \hat{c}_{i}^{\dagger} A_{ij}^{\sigma} \hat{c}_{j},
\end{equation}
where
\begin{align}
A_{ij}^{\sigma} &=\, \frac{1}{2} \left[ \delta_{ij+1} + \delta_{ji+1}
- (P+1) ( \delta_{i1}\delta_{jp} + \delta_{ip}\delta_{j1}  ) \right], \, \notag \\
&\delta_{pp+1} =\, \delta_{p1}, \quad \delta_{p+1p} = \delta_{1p} .
\end{align}
\end{subequations}
It is clear that $A^{\sigma}$ are real and symmetric matrices. 
Hence, their eigenvalues $\Lambda_k$ are real
\begin{align}
\Lambda_{k\sigma} &= \cos(k^{\sigma}), \\ 
\quad k_{\eta}^{\sigma} &= -\pi + \left(2\eta + \frac{\sigma-3}{2}\right) \frac{\pi}{p} , \quad \eta=1,2,\dots, p ,
\end{align}   
and their eigenvectors $\phi_k^{(\sigma)}$ are orthogonal, and can be normalized to give
\begin{align}
\quad \phi_{kj}^{(\sigma)} &= \frac{1}{\sqrt{p}}\,\e^{\i k^{\sigma}j}, \quad 
\sum_{i=1}^{p} \phi_{ki}^{(\sigma)} \phi_{k^{\prime}i}^{(\sigma)\ast} = \delta_{kk^{\prime}}  . 
\end{align}  

The Fourier expansions of $\hat{c}$ and $\hat{c}^{\dagger}$, 
by means of which $A^{\sigma}$ can be diagonalized, are defined via
\begin{align}
\hat{c}_i =\sum_{k \in BZ} \phi^{(\sigma)}_{ki}\hat{\eta}_{k\sigma}, \quad 
\hat{c}_i^{\dagger} =\sum_{k \in BZ} \phi_{ki}^{(\sigma)\ast}\hat{\eta}_{k\sigma}^{\dagger} . \label{c_to_eta}
\end{align}
As can be easily checked the Fourier transforms $\hat{\eta}_{k\sigma}$ 
and $\hat{\eta}^{\dagger}_{k\sigma}$ are also Fermion operators 
\begin{align}
\{\hat{\eta}_{k\sigma}, \hat{\eta}_{k^{\prime}\sigma}^{\dagger}\} = \delta_{kk^{\prime}}, \quad
\{\hat{\eta}_{k\sigma}, \hat{\eta}_{k^{\prime}\sigma}\} = 
\{\hat{\eta}_{k\sigma}^{\dagger}, \hat{\eta}_{k^{\prime}\sigma}^{\dagger}\}  = 0.
\end{align}

Eventually, substituting the expansions (\ref{c_to_eta}) for $\hat{c}$ and $\hat{c}^{\dagger}$
into the expression for $\hat{H}$ (\ref{H_c+i+b}) we end up with
Hamiltonian \eqref{HamiltonianRWA}.

\subsection{Ground state of the $\boldsymbol{XY}$ model} \label{XY_Diagonalization_2}

For our further goals, we also need some of the zero-temperature properties of the $XY$ spin model.
We consider the ground states and energies for the cases of $XY$ model with and without  the external field $B_z$.

\subsubsection{Ground state of the $XY$ model in a strong constant homogeneous field} \label{XY_Diagonalization_21}

For the $XY$ model in the external field $B_z$, we have
\begin{align}
\label{Hamiltonian_XY_Diag_Without_Dikie} 
\hat{H}^{XY + B} &=  \,\, \hbar \omega_1 \, \sum_{i} \hat{S}_{iz}
+ 2J \sum_{i} \left(\hat{S}_{ix} \hat{S}_{i+1x} + \hat{S}_{iy} \hat{S}_{i+1y}\right) \notag \\
&= \,\,  \sum_{\sigma = \pm} \frac{1+\sigma P}{2} \,  \sum_{k \in BZ} \varepsilon_{\sigma}(k) \left(\hat{\eta}_{k\sigma}^{\dagger} \hat{\eta}_{k\sigma} - \frac{1}{2} \right) \, \frac{1+\sigma P}{2},
\end{align}
\begin{equation}
\varepsilon_{\sigma}(k) = 2J\Lambda_{k\sigma} + \hbar \omega_1,
\end{equation}
which is a particular case of (\ref{H_XY_diag}) when no oscillating
field in the $x-y$ -- plane is present ($\Omega = 0, \, \Delta
\rightarrow \omega_1$).
Assuming anti-ferromagnetic $XY$ chain
($J>0$), one distinguishes three cases -- $|\hbar\omega_1| < 2J ,\,
\hbar\omega_1 \leq - 2J$, and $\hbar\omega_1 \geq 2J$.
Henceforth we will consider the regime when $\hbar\omega_1 \geq 2J$.

The values of the Fermi momentum $k_F$ for the free fermion system in
(\ref{Hamiltonian_XY_Diag_Without_Dikie}) are determined by the condition
\begin{align}
\varepsilon_{\sigma}(k_F) = 2J\Lambda_{k\sigma} + \hbar \omega_1 = 0,
\end{align}
from where we find that the Fermi surface consists of the two symmetric points
\begin{align}
k_{\sigma F} = \pm \pi 
\end{align}  
located at the edges of the $BZ$.
Here $\hbar \omega_1$ plays the role of chemical potential.
Since in the regime considered the energy $\varepsilon(k_F)$ 
is nonnegative and determines the minimum of the dispersion relation, i.e.
\begin{align}
\varepsilon(k) \geq \varepsilon(k_F) \geq 0 ,
\end{align}   
the ground state $\vert 0 \rangle$ is characterized by lack of $\eta$--quasi-particles 
\begin{align}
\vert 0 \rangle = \vert 0,0,\dots, 0\rangle,
\end{align}
and hence, ferromagnetic order
\begin{align}
M = \mu \, p, 
\end{align}
where $\mu$ is the projection of the magnetic moment in a single site on the $z$-axis 
and $M$ is the magnetic moment of the chain.
(Recall, that we have assumed $\hat{\mu}_{S} = - 2 \mu \hat{S}$.)

\subsubsection{Ground state of the $XY$ model} \label{XY_Diagonalization_22}

The values of the Fermi momentum $k_F$ in the case of no field present are determined by the condition
\begin{align}
\varepsilon_{\sigma}(k_F) = 2J\Lambda_{k\sigma} = 0.
\end{align}
Now, the Fermi surface consists of the two symmetric points
\begin{align}
k_{\sigma F} = \pm \frac{\pi}{2} .
\end{align}  
The filling numbers of the states for anti-ferromagnetic case ($J>0$) is 
\begin{align}
n_{k\sigma} &= 1 \,\, \text{for} \,\, [-\pi,-k_{\sigma F}] \cup [k_{\sigma F}, \pi] , \\
n_{k\sigma} &= 0  \,\, \text{for} \,\, [-k_{\sigma F},k_{\sigma F}] .
\end{align}
For the ground state energy and ground state, we obtain
\begin{align}
E_{\sigma}(k_F) = 2J \sum_{k \in BZ}\Lambda_k  \hat{\eta}_{k}^{\dagger} \hat{\eta}_{k} , \\
\vert g \rangle = \vert 1,1,\dots, 1, 0,0, \dots, 0, 1,1,\dots, 1\rangle,
\end{align}
respectively.
The system is characterized by anti-ferromagnetic order, so its magnetic moment vanishes,
\begin{align}
M = - \mu \, \left(\sum_{k_{\sigma} \in [-\pi,-k_{\sigma F}]} 1 + \sum_{k_{\sigma} \in [k_{\sigma F}, \pi]} 1 - \sum_{k_{\sigma} \in [-k_{\sigma F}, k_{\sigma F}]} 1 \right) = 0.
\end{align}

In the ferromagnetic case ($J<0$) we have
\begin{align}
E_{\sigma}(k_F) &= 2J \sum_{k \in BZ}\Lambda_k  \hat{\eta}_{k}^{\dagger} \hat{\eta}_{k}= -2|J| \sum_{k \in BZ}\Lambda_k  \hat{\eta}_{k}^{\dagger} \hat{\eta}_{k} , \\
&\vert g' \rangle = \vert 0,0,\dots, 0, 1, \dots, 1, 0,0,\dots, 0\rangle,
\end{align}
and
\begin{align}
M = 0.
\end{align}


\bibliography{jc_xy_arXiv.bib}

\begin{thebibliography}{38}%
\makeatletter
\providecommand \@ifxundefined [1]{%
 \@ifx{#1\undefined}
}%
\providecommand \@ifnum [1]{%
 \ifnum #1\expandafter \@firstoftwo
 \else \expandafter \@secondoftwo
 \fi
}%
\providecommand \@ifx [1]{%
 \ifx #1\expandafter \@firstoftwo
 \else \expandafter \@secondoftwo
 \fi
}%
\providecommand \natexlab [1]{#1}%
\providecommand \enquote  [1]{``#1''}%
\providecommand \bibnamefont  [1]{#1}%
\providecommand \bibfnamefont [1]{#1}%
\providecommand \citenamefont [1]{#1}%
\providecommand \href@noop [0]{\@secondoftwo}%
\providecommand \href [0]{\begingroup \@sanitize@url \@href}%
\providecommand \@href[1]{\@@startlink{#1}\@@href}%
\providecommand \@@href[1]{\endgroup#1\@@endlink}%
\providecommand \@sanitize@url [0]{\catcode `\\12\catcode `\$12\catcode
  `\&12\catcode `\#12\catcode `\^12\catcode `\_12\catcode `\%12\relax}%
\providecommand \@@startlink[1]{}%
\providecommand \@@endlink[0]{}%
\providecommand \url  [0]{\begingroup\@sanitize@url \@url }%
\providecommand \@url [1]{\endgroup\@href {#1}{\urlprefix }}%
\providecommand \urlprefix  [0]{URL }%
\providecommand \Eprint [0]{\href }%
\providecommand \doibase [0]{https://doi.org/}%
\providecommand \selectlanguage [0]{\@gobble}%
\providecommand \bibinfo  [0]{\@secondoftwo}%
\providecommand \bibfield  [0]{\@secondoftwo}%
\providecommand \translation [1]{[#1]}%
\providecommand \BibitemOpen [0]{}%
\providecommand \bibitemStop [0]{}%
\providecommand \bibitemNoStop [0]{.\EOS\space}%
\providecommand \EOS [0]{\spacefactor3000\relax}%
\providecommand \BibitemShut  [1]{\csname bibitem#1\endcsname}%
\let\auto@bib@innerbib\@empty
\bibitem [{\citenamefont {Kirilyuk}\ \emph {et~al.}(2010)\citenamefont
  {Kirilyuk}, \citenamefont {Kimel},\ and\ \citenamefont
  {Rasing}}]{kirilyuk_ultrafast_2010}%
  \BibitemOpen
  \bibfield  {author} {\bibinfo {author} {\bibfnamefont {A.}~\bibnamefont
  {Kirilyuk}}, \bibinfo {author} {\bibfnamefont {A.~V.}\ \bibnamefont
  {Kimel}},\ and\ \bibinfo {author} {\bibfnamefont {T.}~\bibnamefont
  {Rasing}},\ }\bibfield  {title} {\bibinfo {title} {Ultrafast optical
  manipulation of magnetic order},\ }\href
  {https://doi.org/10.1103/RevModPhys.82.2731} {\bibfield  {journal} {\bibinfo
  {journal} {Rev. Mod. Phys.}\ }\textbf {\bibinfo {volume} {82}},\ \bibinfo
  {pages} {2731} (\bibinfo {year} {2010})}\BibitemShut {NoStop}%
\bibitem [{\citenamefont {Noh}\ and\ \citenamefont
  {Angelakis}(2017)}]{noh_quantum_2017}%
  \BibitemOpen
  \bibfield  {author} {\bibinfo {author} {\bibfnamefont {C.}~\bibnamefont
  {Noh}}\ and\ \bibinfo {author} {\bibfnamefont {D.~G.}\ \bibnamefont
  {Angelakis}},\ }\bibfield  {title} {\bibinfo {title} {Quantum simulations and
  many-body physics with light},\ }\href
  {https://doi.org/10.1088/0034-4885/80/1/016401} {\bibfield  {journal}
  {\bibinfo  {journal} {Rep. Prog. Phys.}\ }\textbf {\bibinfo {volume} {80}},\
  \bibinfo {pages} {016401} (\bibinfo {year} {2017})}\BibitemShut {NoStop}%
\bibitem [{\citenamefont {Harder}\ and\ \citenamefont
  {Hu}(2018)}]{harder_cavity_2018}%
  \BibitemOpen
  \bibfield  {author} {\bibinfo {author} {\bibfnamefont {M.}~\bibnamefont
  {Harder}}\ and\ \bibinfo {author} {\bibfnamefont {C.-M.}\ \bibnamefont
  {Hu}},\ }\bibfield  {title} {\bibinfo {title} {{Cavity Spintronics: An Early
  Review of Recent Progress in the Study of Magnon--Photon Level Repulsion}},\
  }in\ \href {https://doi.org/10.1016/bs.ssp.2018.08.001} {\emph {\bibinfo
  {booktitle} {Solid {State} {Phys.}}}},\ Vol.~\bibinfo {volume} {69}\
  (\bibinfo  {publisher} {Elsevier},\ \bibinfo {year} {2018})\ pp.\ \bibinfo
  {pages} {47--121}\BibitemShut {NoStop}%
\bibitem [{\citenamefont {Beaurepaire}\ \emph {et~al.}(1996)\citenamefont
  {Beaurepaire}, \citenamefont {Merle}, \citenamefont {Daunois},\ and\
  \citenamefont {Bigot}}]{Beaurepaire_1995}%
  \BibitemOpen
  \bibfield  {author} {\bibinfo {author} {\bibfnamefont {E.}~\bibnamefont
  {Beaurepaire}}, \bibinfo {author} {\bibfnamefont {J.~C.}\ \bibnamefont
  {Merle}}, \bibinfo {author} {\bibfnamefont {A.}~\bibnamefont {Daunois}},\
  and\ \bibinfo {author} {\bibfnamefont {J.-Y.}\ \bibnamefont {Bigot}},\
  }\bibfield  {title} {\bibinfo {title} {Ultrafast spin dynamics in
  ferromagnetic nickel},\ }\href@noop {} {\bibfield  {journal} {\bibinfo
  {journal} {Phys. Rev. Lett.}\ }\textbf {\bibinfo {volume} {76}},\ \bibinfo
  {pages} {4250} (\bibinfo {year} {1996})}\BibitemShut {NoStop}%
\bibitem [{\citenamefont {Bigot}\ and\ \citenamefont
  {Vomir}(2013)}]{bigot_ultrafast_2013}%
  \BibitemOpen
  \bibfield  {author} {\bibinfo {author} {\bibfnamefont {J.-Y.}\ \bibnamefont
  {Bigot}}\ and\ \bibinfo {author} {\bibfnamefont {M.}~\bibnamefont {Vomir}},\
  }\bibfield  {title} {\bibinfo {title} {Ultrafast magnetization dynamics of
  nanostructures: {Ultrafast} magnetization dynamics of nanostructures},\
  }\href {https://doi.org/10.1002/andp.201200199} {\bibfield  {journal}
  {\bibinfo  {journal} {Ann. Phys.}\ }\textbf {\bibinfo {volume} {525}},\
  \bibinfo {pages} {2} (\bibinfo {year} {2013})}\BibitemShut {NoStop}%
\bibitem [{\citenamefont {Kimel}\ \emph {et~al.}(2007)\citenamefont {Kimel},
  \citenamefont {Kirilyuk},\ and\ \citenamefont
  {Rasing}}]{kimel_femtosecond_2007}%
  \BibitemOpen
  \bibfield  {author} {\bibinfo {author} {\bibfnamefont {A.}~\bibnamefont
  {Kimel}}, \bibinfo {author} {\bibfnamefont {A.}~\bibnamefont {Kirilyuk}},\
  and\ \bibinfo {author} {\bibfnamefont {T.}~\bibnamefont {Rasing}},\
  }\bibfield  {title} {\bibinfo {title} {Femtosecond opto-magnetism: ultrafast
  laser manipulation of magnetic materials},\ }\href
  {https://doi.org/10.1002/lpor.200710022} {\bibfield  {journal} {\bibinfo
  {journal} {Laser Photonics Rev.}\ }\textbf {\bibinfo {volume} {1}},\ \bibinfo
  {pages} {275} (\bibinfo {year} {2007})}\BibitemShut {NoStop}%
\bibitem [{\citenamefont {Zhang}\ \emph {et~al.}(2016)\citenamefont {Zhang},
  \citenamefont {Latta}, \citenamefont {Babyak}, \citenamefont {Bai},\ and\
  \citenamefont {George}}]{zhang_all-optical_2016}%
  \BibitemOpen
  \bibfield  {author} {\bibinfo {author} {\bibfnamefont {G.~P.}\ \bibnamefont
  {Zhang}}, \bibinfo {author} {\bibfnamefont {T.}~\bibnamefont {Latta}},
  \bibinfo {author} {\bibfnamefont {Z.}~\bibnamefont {Babyak}}, \bibinfo
  {author} {\bibfnamefont {Y.~H.}\ \bibnamefont {Bai}},\ and\ \bibinfo {author}
  {\bibfnamefont {T.~F.}\ \bibnamefont {George}},\ }\bibfield  {title}
  {\bibinfo {title} {{All-optical spin switching: A new frontier in
  femtomagnetism -- A short review and a simple theory}},\ }\href
  {https://doi.org/10.1142/S0217984916300052} {\bibfield  {journal} {\bibinfo
  {journal} {Mod. Phys. Lett. B}\ }\textbf {\bibinfo {volume} {30}},\ \bibinfo
  {pages} {16300052} (\bibinfo {year} {2016})}\BibitemShut {NoStop}%
\bibitem [{\citenamefont {Zhang}\ \emph {et~al.}(2014)\citenamefont {Zhang},
  \citenamefont {Gu},\ and\ \citenamefont {Wu}}]{zhang_ultrafast_2014}%
  \BibitemOpen
  \bibfield  {author} {\bibinfo {author} {\bibfnamefont {G.~P.}\ \bibnamefont
  {Zhang}}, \bibinfo {author} {\bibfnamefont {M.}~\bibnamefont {Gu}},\ and\
  \bibinfo {author} {\bibfnamefont {X.~S.}\ \bibnamefont {Wu}},\ }\bibfield
  {title} {\bibinfo {title} {Ultrafast reduction in exchange interaction by a
  laser pulse: alternative path to femtomagnetism},\ }\href
  {https://doi.org/10.1088/0953-8984/26/37/376001} {\bibfield  {journal}
  {\bibinfo  {journal} {J. Phys.: Condens. Matter}\ }\textbf {\bibinfo {volume}
  {26}},\ \bibinfo {pages} {376001} (\bibinfo {year} {2014})}\BibitemShut
  {NoStop}%
\bibitem [{\citenamefont {Bossini}\ \emph {et~al.}(2014)\citenamefont
  {Bossini}, \citenamefont {Kalashnikova}, \citenamefont {Pisarev},
  \citenamefont {Rasing},\ and\ \citenamefont
  {Kimel}}]{bossini_controlling_2014}%
  \BibitemOpen
  \bibfield  {author} {\bibinfo {author} {\bibfnamefont {D.}~\bibnamefont
  {Bossini}}, \bibinfo {author} {\bibfnamefont {A.~M.}\ \bibnamefont
  {Kalashnikova}}, \bibinfo {author} {\bibfnamefont {R.~V.}\ \bibnamefont
  {Pisarev}}, \bibinfo {author} {\bibfnamefont {T.}~\bibnamefont {Rasing}},\
  and\ \bibinfo {author} {\bibfnamefont {A.~V.}\ \bibnamefont {Kimel}},\
  }\bibfield  {title} {\bibinfo {title} {Controlling coherent and incoherent
  spin dynamics by steering the photoinduced energy flow},\ }\href
  {https://doi.org/10.1103/PhysRevB.89.060405} {\bibfield  {journal} {\bibinfo
  {journal} {Phys. Rev. B}\ }\textbf {\bibinfo {volume} {89}},\ \bibinfo
  {pages} {060405(R)} (\bibinfo {year} {2014})}\BibitemShut {NoStop}%
\bibitem [{\citenamefont {Hansteen}\ \emph {et~al.}(2005)\citenamefont
  {Hansteen}, \citenamefont {Kimel}, \citenamefont {Kirilyuk},\ and\
  \citenamefont {Rasing}}]{hansteen_femtosecond_2005}%
  \BibitemOpen
  \bibfield  {author} {\bibinfo {author} {\bibfnamefont {F.}~\bibnamefont
  {Hansteen}}, \bibinfo {author} {\bibfnamefont {A.}~\bibnamefont {Kimel}},
  \bibinfo {author} {\bibfnamefont {A.}~\bibnamefont {Kirilyuk}},\ and\
  \bibinfo {author} {\bibfnamefont {T.}~\bibnamefont {Rasing}},\ }\bibfield
  {title} {\bibinfo {title} {{Femtosecond Photomagnetic Switching of Spins in
  Ferrimagnetic Garnet Films}},\ }\href
  {https://doi.org/10.1103/PhysRevLett.95.047402} {\bibfield  {journal}
  {\bibinfo  {journal} {Phys. Rev. Lett.}\ }\textbf {\bibinfo {volume} {95}},\
  \bibinfo {pages} {047402} (\bibinfo {year} {2005})}\BibitemShut {NoStop}%
\bibitem [{\citenamefont {Brune}(2006)}]{brune_assembly_2006}%
  \BibitemOpen
  \bibfield  {author} {\bibinfo {author} {\bibfnamefont {H.}~\bibnamefont
  {Brune}},\ }\bibfield  {title} {\bibinfo {title} {Assembly and {Probing} of
  {Spin} {Chains} of {Finite} {Size}},\ }\href
  {https://doi.org/10.1126/science.1127387} {\bibfield  {journal} {\bibinfo
  {journal} {Science}\ }\textbf {\bibinfo {volume} {312}},\ \bibinfo {pages}
  {1005} (\bibinfo {year} {2006})}\BibitemShut {NoStop}%
\bibitem [{\citenamefont {Chudnovsky}\ and\ \citenamefont
  {Garanin}(2002)}]{Chudnovsky2002}%
  \BibitemOpen
  \bibfield  {author} {\bibinfo {author} {\bibfnamefont {E.~M.}\ \bibnamefont
  {Chudnovsky}}\ and\ \bibinfo {author} {\bibfnamefont {D.~A.}\ \bibnamefont
  {Garanin}},\ }\bibfield  {title} {\bibinfo {title} {Superradiance from
  crystals of molecular nanomagnets},\ }\href
  {https://doi.org/10.1103/PhysRevLett.89.157201} {\bibfield  {journal}
  {\bibinfo  {journal} {Phys. Rev. Lett.}\ }\textbf {\bibinfo {volume} {89}},\
  \bibinfo {pages} {157201} (\bibinfo {year} {2002})}\BibitemShut {NoStop}%
\bibitem [{\citenamefont {Soykal}\ and\ \citenamefont
  {Flatt\'{e}}(2010)}]{Soykal2010}%
  \BibitemOpen
  \bibfield  {author} {\bibinfo {author} {\bibfnamefont {{\"{O}}.~O.}\
  \bibnamefont {Soykal}}\ and\ \bibinfo {author} {\bibfnamefont {M.~E.}\
  \bibnamefont {Flatt\'{e}}},\ }\bibfield  {title} {\bibinfo {title} {{Strong
  Field Interactions between a Nanomagnet and a Photonic Cavity}},\ }\href
  {https://doi.org/10.1103/PhysRevLett.104.077202} {\bibfield  {journal}
  {\bibinfo  {journal} {Phys. Rev. Lett.}\ }\textbf {\bibinfo {volume} {104}},\
  \bibinfo {pages} {077202} (\bibinfo {year} {2010})}\BibitemShut {NoStop}%
\bibitem [{\citenamefont {Jaynes}\ and\ \citenamefont
  {Cummings}(1963)}]{Jaynes1963}%
  \BibitemOpen
  \bibfield  {author} {\bibinfo {author} {\bibfnamefont {E.}~\bibnamefont
  {Jaynes}}\ and\ \bibinfo {author} {\bibfnamefont {F.}~\bibnamefont
  {Cummings}},\ }\bibfield  {title} {\bibinfo {title} {Comparison of quantum
  and semiclassical radiation theories with application to the beam maser},\
  }\href {https://doi.org/10.1109/PROC.1963.1664} {\bibfield  {journal}
  {\bibinfo  {journal} {Proc. IEEE}\ }\textbf {\bibinfo {volume} {51}},\
  \bibinfo {pages} {89} (\bibinfo {year} {1963})}\BibitemShut {NoStop}%
\bibitem [{\citenamefont {Tavis}\ and\ \citenamefont
  {Cummings}(1968)}]{Tavis1968}%
  \BibitemOpen
  \bibfield  {author} {\bibinfo {author} {\bibfnamefont {M.}~\bibnamefont
  {Tavis}}\ and\ \bibinfo {author} {\bibfnamefont {F.~W.}\ \bibnamefont
  {Cummings}},\ }\bibfield  {title} {\bibinfo {title} {{Exact Solution for an
  $N$-Molecule-Radiation-Field Hamiltonian}},\ }\href
  {https://doi.org/10.1103/PhysRev.170.379} {\bibfield  {journal} {\bibinfo
  {journal} {Phys. Rev.}\ }\textbf {\bibinfo {volume} {170}},\ \bibinfo {pages}
  {379} (\bibinfo {year} {1968})}\BibitemShut {NoStop}%
\bibitem [{\citenamefont {Dicke}(1954)}]{Dicke1954}%
  \BibitemOpen
  \bibfield  {author} {\bibinfo {author} {\bibfnamefont {R.~H.}\ \bibnamefont
  {Dicke}},\ }\bibfield  {title} {\bibinfo {title} {{Coherence in Spontaneous
  Radiation Processes}},\ }\href {https://doi.org/10.1103/PhysRev.93.99}
  {\bibfield  {journal} {\bibinfo  {journal} {Phys. Rev.}\ }\textbf {\bibinfo
  {volume} {93}},\ \bibinfo {pages} {99} (\bibinfo {year} {1954})}\BibitemShut
  {NoStop}%
\bibitem [{\citenamefont {Kirton}\ \emph {et~al.}(2019)\citenamefont {Kirton},
  \citenamefont {Roses}, \citenamefont {Keeling},\ and\ \citenamefont
  {Dalla~Torre}}]{kirton_introduction_2019}%
  \BibitemOpen
  \bibfield  {author} {\bibinfo {author} {\bibfnamefont {P.}~\bibnamefont
  {Kirton}}, \bibinfo {author} {\bibfnamefont {M.~M.}\ \bibnamefont {Roses}},
  \bibinfo {author} {\bibfnamefont {J.}~\bibnamefont {Keeling}},\ and\ \bibinfo
  {author} {\bibfnamefont {E.~G.}\ \bibnamefont {Dalla~Torre}},\ }\bibfield
  {title} {\bibinfo {title} {Introduction to the {Dicke} {Model}: {From}
  {Equilibrium} to {Nonequilibrium}, and \textit{{Vice} {Versa}}},\ }\href
  {https://doi.org/10.1002/qute.201800043} {\bibfield  {journal} {\bibinfo
  {journal} {Adv. Quantum Technol.}\ }\textbf {\bibinfo {volume} {2}},\
  \bibinfo {pages} {1800043} (\bibinfo {year} {2019})}\BibitemShut {NoStop}%
\bibitem [{\citenamefont {Higgins}\ \emph {et~al.}(2014)\citenamefont
  {Higgins}, \citenamefont {Benjamin}, \citenamefont {Stace}, \citenamefont
  {Milburn}, \citenamefont {Lovett},\ and\ \citenamefont
  {Gauger}}]{Higgins_superabsorption_2014}%
  \BibitemOpen
  \bibfield  {author} {\bibinfo {author} {\bibfnamefont {K.~D.~B.}\
  \bibnamefont {Higgins}}, \bibinfo {author} {\bibfnamefont {S.~C.}\
  \bibnamefont {Benjamin}}, \bibinfo {author} {\bibfnamefont {T.~M.}\
  \bibnamefont {Stace}}, \bibinfo {author} {\bibfnamefont {G.~J.}\ \bibnamefont
  {Milburn}}, \bibinfo {author} {\bibfnamefont {B.~W.}\ \bibnamefont
  {Lovett}},\ and\ \bibinfo {author} {\bibfnamefont {E.~M.}\ \bibnamefont
  {Gauger}},\ }\bibfield  {title} {\bibinfo {title} {Superabsorption of light
  via quantum engineering},\ }\href {https://doi.org/10.1038/ncomms5705}
  {\bibfield  {journal} {\bibinfo  {journal} {Nat. Commun.}\ }\textbf {\bibinfo
  {volume} {5}},\ \bibinfo {pages} {4705} (\bibinfo {year} {2014})}\BibitemShut
  {NoStop}%
\bibitem [{\citenamefont {Grinberg}(2010)}]{grinberg_beyond_2010}%
  \BibitemOpen
  \bibfield  {author} {\bibinfo {author} {\bibfnamefont {H.}~\bibnamefont
  {Grinberg}},\ }\bibfield  {title} {\bibinfo {title} {{Beyond the rotating
  wave approximation. An intensity dependent nonlinear coupling model in
  two-level systems}},\ }\href {https://doi.org/10.1016/j.physleta.2010.01.047}
  {\bibfield  {journal} {\bibinfo  {journal} {Phys. Lett. A}\ }\textbf
  {\bibinfo {volume} {374}},\ \bibinfo {pages} {1481} (\bibinfo {year}
  {2010})}\BibitemShut {NoStop}%
\bibitem [{\citenamefont {Agarwal}\ \emph {et~al.}(2012)\citenamefont
  {Agarwal}, \citenamefont {Rafsanjani},\ and\ \citenamefont
  {Eberly}}]{Agarwal2012}%
  \BibitemOpen
  \bibfield  {author} {\bibinfo {author} {\bibfnamefont {S.}~\bibnamefont
  {Agarwal}}, \bibinfo {author} {\bibfnamefont {S.~M.~H.}\ \bibnamefont
  {Rafsanjani}},\ and\ \bibinfo {author} {\bibfnamefont {J.~H.}\ \bibnamefont
  {Eberly}},\ }\bibfield  {title} {\bibinfo {title} {Tavis-{Cummings} model
  beyond the rotating wave approximation: {Quasidegenerate} qubits},\ }\href
  {https://doi.org/10.1103/PhysRevA.85.043815} {\bibfield  {journal} {\bibinfo
  {journal} {Phys. Rev. A}\ }\textbf {\bibinfo {volume} {85}},\ \bibinfo
  {pages} {043815} (\bibinfo {year} {2012})}\BibitemShut {NoStop}%
\bibitem [{\citenamefont {Agarwal}\ \emph {et~al.}(1997)\citenamefont
  {Agarwal}, \citenamefont {Puri},\ and\ \citenamefont
  {Singh}}]{agarwal_atomic_1997}%
  \BibitemOpen
  \bibfield  {author} {\bibinfo {author} {\bibfnamefont {G.~S.}\ \bibnamefont
  {Agarwal}}, \bibinfo {author} {\bibfnamefont {R.~R.}\ \bibnamefont {Puri}},\
  and\ \bibinfo {author} {\bibfnamefont {R.~P.}\ \bibnamefont {Singh}},\
  }\bibfield  {title} {\bibinfo {title} {Atomic {Schr\"{o}dinger} cat states},\
  }\href {https://doi.org/10.1103/PhysRevA.56.2249} {\bibfield  {journal}
  {\bibinfo  {journal} {Phys. Rev. A}\ }\textbf {\bibinfo {volume} {56}},\
  \bibinfo {pages} {2249} (\bibinfo {year} {1997})}\BibitemShut {NoStop}%
\bibitem [{\citenamefont {Klimov}\ and\ \citenamefont
  {Saavedra}(1998)}]{Klimov1998}%
  \BibitemOpen
  \bibfield  {author} {\bibinfo {author} {\bibfnamefont {A.}~\bibnamefont
  {Klimov}}\ and\ \bibinfo {author} {\bibfnamefont {C.}~\bibnamefont
  {Saavedra}},\ }\bibfield  {title} {\bibinfo {title} {{The Dicke model
  dynamics in a high detuning limit}},\ }\href
  {https://doi.org/10.1016/S0375-9601(98)00529-5} {\bibfield  {journal}
  {\bibinfo  {journal} {Phys. Lett. A}\ }\textbf {\bibinfo {volume} {247}},\
  \bibinfo {pages} {14} (\bibinfo {year} {1998})}\BibitemShut {NoStop}%
\bibitem [{\citenamefont {Str\"{a}ter}\ \emph {et~al.}(2012)\citenamefont
  {Str\"{a}ter}, \citenamefont {Tsyplyatyev},\ and\ \citenamefont
  {Faribault}}]{Strater2012}%
  \BibitemOpen
  \bibfield  {author} {\bibinfo {author} {\bibfnamefont {C.}~\bibnamefont
  {Str\"{a}ter}}, \bibinfo {author} {\bibfnamefont {O.}~\bibnamefont
  {Tsyplyatyev}},\ and\ \bibinfo {author} {\bibfnamefont {A.}~\bibnamefont
  {Faribault}},\ }\bibfield  {title} {\bibinfo {title} {Nonequilibrum dynamics
  in the strongly excited inhomogeneous {Dicke} model},\ }\href
  {https://doi.org/10.1103/PhysRevB.86.195101} {\bibfield  {journal} {\bibinfo
  {journal} {Phys. Rev. B}\ }\textbf {\bibinfo {volume} {86}},\ \bibinfo
  {pages} {195101} (\bibinfo {year} {2012})}\BibitemShut {NoStop}%
\bibitem [{\citenamefont {Tokihiro}\ \emph {et~al.}(1993)\citenamefont
  {Tokihiro}, \citenamefont {Manabe},\ and\ \citenamefont
  {Hanamura}}]{Tokihiro1993}%
  \BibitemOpen
  \bibfield  {author} {\bibinfo {author} {\bibfnamefont {T.}~\bibnamefont
  {Tokihiro}}, \bibinfo {author} {\bibfnamefont {Y.}~\bibnamefont {Manabe}},\
  and\ \bibinfo {author} {\bibfnamefont {E.}~\bibnamefont {Hanamura}},\
  }\bibfield  {title} {\bibinfo {title} {Superradiance of {Frenkel} excitons in
  linear systems},\ }\href {https://doi.org/10.1103/PhysRevB.47.2019}
  {\bibfield  {journal} {\bibinfo  {journal} {Phys. Rev. B}\ }\textbf {\bibinfo
  {volume} {47}},\ \bibinfo {pages} {2019} (\bibinfo {year}
  {1993})}\BibitemShut {NoStop}%
\bibitem [{\citenamefont {Wu}\ \emph {et~al.}(2016)\citenamefont {Wu},
  \citenamefont {Feist},\ and\ \citenamefont {Garcia-Vidal}}]{Wu2016}%
  \BibitemOpen
  \bibfield  {author} {\bibinfo {author} {\bibfnamefont {N.}~\bibnamefont
  {Wu}}, \bibinfo {author} {\bibfnamefont {J.}~\bibnamefont {Feist}},\ and\
  \bibinfo {author} {\bibfnamefont {F.~J.}\ \bibnamefont {Garcia-Vidal}},\
  }\bibfield  {title} {\bibinfo {title} {When polarons meet polaritons:
  {Exciton}-vibration interactions in organic molecules strongly coupled to
  confined light fields},\ }\href {https://doi.org/10.1103/PhysRevB.94.195409}
  {\bibfield  {journal} {\bibinfo  {journal} {Phys. Rev. B}\ }\textbf {\bibinfo
  {volume} {94}},\ \bibinfo {pages} {195409} (\bibinfo {year}
  {2016})}\BibitemShut {NoStop}%
\bibitem [{\citenamefont {Wu}(2018)}]{Wu2018}%
  \BibitemOpen
  \bibfield  {author} {\bibinfo {author} {\bibfnamefont {N.}~\bibnamefont
  {Wu}},\ }\bibfield  {title} {\bibinfo {title} {{Determinant representations
  of spin-operator matrix elements in the {XX} spin chain and their
  applications}},\ }\href {https://doi.org/10.1103/PhysRevB.97.014301}
  {\bibfield  {journal} {\bibinfo  {journal} {Phys. Rev. B}\ }\textbf {\bibinfo
  {volume} {97}},\ \bibinfo {pages} {014301} (\bibinfo {year}
  {2018})}\BibitemShut {NoStop}%
\bibitem [{\citenamefont {Tonchev}\ \emph
  {et~al.}(2016{\natexlab{a}})\citenamefont {Tonchev}, \citenamefont {Donkov},\
  and\ \citenamefont {Chamati}}]{tonchev_interaction_2016}%
  \BibitemOpen
  \bibfield  {author} {\bibinfo {author} {\bibfnamefont {H.}~\bibnamefont
  {Tonchev}}, \bibinfo {author} {\bibfnamefont {A.~A.}\ \bibnamefont
  {Donkov}},\ and\ \bibinfo {author} {\bibfnamefont {H.}~\bibnamefont
  {Chamati}},\ }\bibfield  {title} {\bibinfo {title} {{Interaction of a single
  mode field cavity with the 1D {XY} model: {Energy} spectrum}},\ }\href
  {https://doi.org/10.1088/1742-6596/682/1/012032} {\bibfield  {journal}
  {\bibinfo  {journal} {J. Phys. Conf. Ser.}\ }\textbf {\bibinfo {volume}
  {682}},\ \bibinfo {pages} {012032} (\bibinfo {year}
  {2016}{\natexlab{a}})}\BibitemShut {NoStop}%
\bibitem [{\citenamefont {Tonchev}\ \emph
  {et~al.}(2016{\natexlab{b}})\citenamefont {Tonchev}, \citenamefont {Donkov},\
  and\ \citenamefont {Chamati}}]{tonchev_energy_2016}%
  \BibitemOpen
  \bibfield  {author} {\bibinfo {author} {\bibfnamefont {H.}~\bibnamefont
  {Tonchev}}, \bibinfo {author} {\bibfnamefont {A.~A.}\ \bibnamefont
  {Donkov}},\ and\ \bibinfo {author} {\bibfnamefont {H.}~\bibnamefont
  {Chamati}},\ }\bibfield  {title} {\bibinfo {title} {{Energy spectra of a
  spin-$\tfrac12$ {XY} spin molecule interacting with a single mode field
  cavity: {Numerical} study}},\ }\href
  {https://doi.org/10.1088/1742-6596/764/1/012017} {\bibfield  {journal}
  {\bibinfo  {journal} {J. Phys. Conf. Ser.}\ }\textbf {\bibinfo {volume}
  {764}},\ \bibinfo {pages} {012017} (\bibinfo {year}
  {2016}{\natexlab{b}})}\BibitemShut {NoStop}%
\bibitem [{\citenamefont {Tonchev}\ \emph {et~al.}(2019)\citenamefont
  {Tonchev}, \citenamefont {Donkov},\ and\ \citenamefont
  {Chamati}}]{tonchev_energy_2019}%
  \BibitemOpen
  \bibfield  {author} {\bibinfo {author} {\bibfnamefont {H.}~\bibnamefont
  {Tonchev}}, \bibinfo {author} {\bibfnamefont {A.~A.}\ \bibnamefont
  {Donkov}},\ and\ \bibinfo {author} {\bibfnamefont {H.}~\bibnamefont
  {Chamati}},\ }\bibfield  {title} {\bibinfo {title} {{Energy spectra of a
  spin-$\tfrac12$ {XY} spin molecule interacting with a single mode field
  cavity}},\ }\href {https://doi.org/10.1088/1742-6596/1186/1/012021}
  {\bibfield  {journal} {\bibinfo  {journal} {J. Phys. Conf. Ser.}\ }\textbf
  {\bibinfo {volume} {1186}},\ \bibinfo {pages} {012021} (\bibinfo {year}
  {2019})}\BibitemShut {NoStop}%
\bibitem [{\citenamefont {Lieb}\ \emph {et~al.}(1961)\citenamefont {Lieb},
  \citenamefont {Schultz},\ and\ \citenamefont {Mattis}}]{Lieb1961a}%
  \BibitemOpen
  \bibfield  {author} {\bibinfo {author} {\bibfnamefont {E.}~\bibnamefont
  {Lieb}}, \bibinfo {author} {\bibfnamefont {T.}~\bibnamefont {Schultz}},\ and\
  \bibinfo {author} {\bibfnamefont {D.}~\bibnamefont {Mattis}},\ }\bibfield
  {title} {\bibinfo {title} {Two soluble models of an antiferromagnetic
  chain},\ }\href {https://doi.org/10.1016/0003-4916(61)90115-4} {\bibfield
  {journal} {\bibinfo  {journal} {Ann. Phys.}\ }\textbf {\bibinfo {volume}
  {16}},\ \bibinfo {pages} {407} (\bibinfo {year} {1961})}\BibitemShut
  {NoStop}%
\bibitem [{\citenamefont {Lieb}\ and\ \citenamefont
  {Mattis}(1966)}]{E.Lieb1966b}%
  \BibitemOpen
  \bibfield  {author} {\bibinfo {author} {\bibfnamefont {E.~H.}\ \bibnamefont
  {Lieb}}\ and\ \bibinfo {author} {\bibfnamefont {D.~C.}\ \bibnamefont
  {Mattis}},\ }\href@noop {} {\emph {\bibinfo {title} {{Mathematical physics in
  one dimension: Exactly soluble models of interacting particles.}}}}\
  (\bibinfo  {publisher} {Academic Press},\ \bibinfo {address} {New York},\
  \bibinfo {year} {1966})\BibitemShut {NoStop}%
\bibitem [{\citenamefont {De~Pasquale}\ \emph {et~al.}(2008)\citenamefont
  {De~Pasquale}, \citenamefont {Costantini}, \citenamefont {Facchi},
  \citenamefont {Florio}, \citenamefont {Pascazio},\ and\ \citenamefont
  {Yuasa}}]{DePasquale2008}%
  \BibitemOpen
  \bibfield  {author} {\bibinfo {author} {\bibfnamefont {A.}~\bibnamefont
  {De~Pasquale}}, \bibinfo {author} {\bibfnamefont {G.}~\bibnamefont
  {Costantini}}, \bibinfo {author} {\bibfnamefont {P.}~\bibnamefont {Facchi}},
  \bibinfo {author} {\bibfnamefont {G.}~\bibnamefont {Florio}}, \bibinfo
  {author} {\bibfnamefont {S.}~\bibnamefont {Pascazio}},\ and\ \bibinfo
  {author} {\bibfnamefont {K.}~\bibnamefont {Yuasa}},\ }\bibfield  {title}
  {\bibinfo {title} {{XX} model on the circle},\ }\href
  {https://doi.org/10.1140/epjst/e2008-00716-9} {\bibfield  {journal} {\bibinfo
   {journal} {Eur. Phys. J. Spec. Top.}\ }\textbf {\bibinfo {volume} {160}},\
  \bibinfo {pages} {127} (\bibinfo {year} {2008})}\BibitemShut {NoStop}%
\bibitem [{\citenamefont {Klimov}\ \emph {et~al.}(2002)\citenamefont {Klimov},
  \citenamefont {S\'{a}nchez-Soto}, \citenamefont {Navarro},\ and\
  \citenamefont {Yustas}}]{Klimov_2002}%
  \BibitemOpen
  \bibfield  {author} {\bibinfo {author} {\bibfnamefont {A.~B.}\ \bibnamefont
  {Klimov}}, \bibinfo {author} {\bibfnamefont {L.~L.}\ \bibnamefont
  {S\'{a}nchez-Soto}}, \bibinfo {author} {\bibfnamefont {A.}~\bibnamefont
  {Navarro}},\ and\ \bibinfo {author} {\bibfnamefont {E.~C.}\ \bibnamefont
  {Yustas}},\ }\bibfield  {title} {\bibinfo {title} {Effective hamiltonians in
  quantum optics: a systematic approach},\ }\href
  {https://doi.org/10.1080/09500340210134675} {\bibfield  {journal} {\bibinfo
  {journal} {J. Mod. Opt.}\ }\textbf {\bibinfo {volume} {49}},\ \bibinfo
  {pages} {2211} (\bibinfo {year} {2002})}\BibitemShut {NoStop}%
\bibitem [{\citenamefont {Jelley}(1936)}]{Jelley1936}%
  \BibitemOpen
  \bibfield  {author} {\bibinfo {author} {\bibfnamefont {E.~E.}\ \bibnamefont
  {Jelley}},\ }\href@noop {} {\bibfield  {journal} {\bibinfo  {journal}
  {Nature}\ }\textbf {\bibinfo {volume} {138}},\ \bibinfo {pages} {1009}
  (\bibinfo {year} {1936})}\BibitemShut {NoStop}%
\bibitem [{\citenamefont {Sheibe}(1936)}]{Sheibe1936}%
  \BibitemOpen
  \bibfield  {author} {\bibinfo {author} {\bibfnamefont {G.}~\bibnamefont
  {Sheibe}},\ }\href@noop {} {\bibfield  {journal} {\bibinfo  {journal} {Angew.
  Chem}\ }\textbf {\bibinfo {volume} {49}},\ \bibinfo {pages} {563} (\bibinfo
  {year} {1936})}\BibitemShut {NoStop}%
\bibitem [{\citenamefont {Kobayasshi}(1996)}]{Kobayashi1996}%
  \BibitemOpen
  \bibinfo {editor} {\bibfnamefont {T.}~\bibnamefont {Kobayasshi}},\ ed.,\
  \href@noop {} {\emph {\bibinfo {title} {J-Aggregates}}},\ Vol.~\bibinfo
  {volume} {1}\ (\bibinfo  {publisher} {World Scientific},\ \bibinfo {address}
  {Singapore},\ \bibinfo {year} {1996})\BibitemShut {NoStop}%
\bibitem [{\citenamefont {Kobayasshi}(2012)}]{Kobayashi2012}%
  \BibitemOpen
  \bibinfo {editor} {\bibfnamefont {T.}~\bibnamefont {Kobayasshi}},\ ed.,\
  \href@noop {} {\emph {\bibinfo {title} {J-Aggregates}}},\ Vol.~\bibinfo
  {volume} {2}\ (\bibinfo  {publisher} {World Scientific},\ \bibinfo {address}
  {Singapore},\ \bibinfo {year} {2012})\BibitemShut {NoStop}%
\bibitem [{\citenamefont {Puri}(2001)}]{Puri2001}%
  \BibitemOpen
  \bibfield  {author} {\bibinfo {author} {\bibfnamefont {R.~R.}\ \bibnamefont
  {Puri}},\ }\href {https://doi.org/10.1007/978-3-540-44953-9} {\emph {\bibinfo
  {title} {Mathematical {Methods} of {Quantum} {Optics}}}},\ edited by\
  \bibinfo {editor} {\bibfnamefont {W.~T.}\ \bibnamefont {Rhodes}},\ \bibinfo
  {series} {Springer {Series} in {Optical} {Sciences}}, Vol.~\bibinfo {volume}
  {79}\ (\bibinfo  {publisher} {Springer Berlin Heidelberg},\ \bibinfo
  {address} {Berlin, Heidelberg},\ \bibinfo {year} {2001})\BibitemShut
  {NoStop}%
\end{thebibliography}%
\end{document}